\documentclass[12pt, draftclsnofoot, onecolumn]{IEEEtran}

\usepackage[acronym]{glossaries}
\newacronym{AP}{AP}{analog precoder}
\newacronym{AWGN}{AWGN}{additive white gaussian noise}
\newacronym{BS}{BS}{base station}
\newacronym{MIMO}{MIMO}{multiple-input multiple-output}
\newacronym{MISO}{MISO}{multiple-input single-output}
\newacronym{RF}{RF}{radio-frequency}
\newacronym{CL}{CL}{convolutional layer}
\newacronym{FDD}{FDD}{frequency division duplex}
\newacronym{CSI}{CSI}{channel state information}
\newacronym{DNN}{DNN}{deep neural network}
\newacronym{DP}{DP}{digital precoder}
\newacronym{HSHO}{HSHO}{Hybrid Structured Heuristic Optimization}
\newacronym{DL}{DL}{deep learning}
\newacronym{CNN}{CNN}{convolution neural network}
\newacronym{FDP}{FDP}{fully digital precoder}
\newacronym{OFDM}{OFDM}{orthogonal frequency division multiplexing}
\newacronym{OMP}{OMP}{Orthogonal Matching Pursuit}
\newacronym{FL}{FL}{fully-connected layer}
\newacronym{HBF}{HBF}{hybrid beamforming}
\newacronym{IA}{IA}{initial access}
\newacronym{mm-Wave}{mm-Wave}{millimeter wave}
\newacronym{SINR}{SINR}{signal-to-interference-noise ratio}
\newacronym{SNR}{SNR}{signal-to-noise ratio}
\newacronym{SS}{SS}{synchronization signal}
\newacronym{SSB}{SSB}{synchronization signal burst}
\newacronym{RSSI}{RSSI}{received signal strength indicators}
\newacronym{PZF}{PZF}{phase zero forcing}
\newacronym{ZF}{ZF}{zero forcing}
\newacronym{IoT}{IoT}{internet-of-things}

\newif\ifDeepMIMOModel
\DeepMIMOModeltrue

\newif\ifSimpleNParamEq
\SimpleNParamEqtrue

\usepackage{lettrine}

\usepackage{ifpdf}

\ifCLASSINFOpdf
  \usepackage[pdftex]{graphicx}
  \graphicspath{{../pdf/}{../jpeg/}}
  \DeclareGraphicsExtensions{.pdf,.jpeg,.png}
\else
  \usepackage[dvips]{graphicx}
  \graphicspath{{../eps/}}
  \DeclareGraphicsExtensions{.eps}
\fi

\usepackage{cite}
\usepackage{algorithmic}
\usepackage{amsthm}
\usepackage{amssymb}
\usepackage{balance}
\usepackage{eqparbox}
\usepackage{multirow}
\usepackage[ruled,vlined]{algorithm2e}
\usepackage{amsmath}

\usepackage{mathtools, nccmath}
\DeclarePairedDelimiter{\nint}\lfloor\rceil

\usepackage{booktabs, siunitx}
\usepackage[svgnames,table]{xcolor}

\usepackage{graphicx}
\usepackage[scaled]{DejaVuSansMono}
\usepackage[T1]{fontenc}

\usepackage{color}
\usepackage{relsize}
\usepackage{mathtools}

\newcommand{\bs}[1]{\boldsymbol{#1}}

\newcommand{\mb}[1]{\mathbf{#1}}

\DeclareMathOperator*{\argmax}{arg\;max}

\renewcommand{\Pr}{\mathbb{P}} 
\SetKwInput{KwInput}{Input}                
\SetKwInput{KwOutputr}{Output Regression}              
\SetKwInput{KwOutputc}{Output Classification}              

\newcommand{\bseq}{\begin{subequations}}
\newcommand{\eseq}{\end{subequations}}
\newcommand{\baln}{\begin{align}}
\newcommand{\ealn}{\end{align}}
\newcommand{\balnd}{\begin{aligned}}
\newcommand{\ealnd}{\end{aligned}}
\newcommand{\beq}{\begin{equation}}
\newcommand{\eeq}{\end{equation}}
\newcommand{\beqn}{\begin{eqnarray}}
\newcommand{\eeqn}{\end{eqnarray}}
\newcommand{\beqno}{\begin{eqnarray*}}
\newcommand{\eeqno}{\end{eqnarray*}}
\newcommand{\bma}{\begin{displaymath}}
\newcommand{\ema}{\end{displaymath}}
\newcommand{\bnu}{\begin{enumerate}}
\newcommand{\enu}{\end{enumerate}}
\newcommand{\bce}{\begin{center}}
\newcommand{\ece}{\end{center}}
\newcommand{\btb}{\begin{tabular}}
\newcommand{\etb}{\end{tabular}}
\newcommand{\ba}{\begin{array}}
\newcommand{\ea}{\end{array}}

\begin{document}
\title{Unsupervised Deep Learning for Massive MIMO Hybrid Beamforming}
\author{Hamed Hojatian, J\'{e}r\'{e}my Nadal, Jean-Fran\c{c}ois Frigon, and Fran\c{c}ois Leduc-Primeau \\
{\small\'{E}cole Polytechnique de Montr\'{e}al, Montreal, Quebec, Canada, H3T 1J4 \\ 
Emails: \{hamed.hojatian, jeremy.nadal, j-f.frigon, francois.leduc-primeau\}@polymtl.ca} }

\maketitle

\begin{abstract}
Hybrid beamforming is a promising technique to reduce the complexity and cost of massive multiple-input multiple-output (MIMO) systems while providing high data rate. However, the hybrid precoder design is a challenging task requiring channel state information (CSI) feedback and solving a complex optimization problem. This paper proposes a novel RSSI-based unsupervised deep learning method to design the hybrid beamforming in massive MIMO systems. Furthermore, we propose i) a method to design the synchronization signal (SS) in initial access (IA); and ii) a method to design the codebook for the analog precoder. We also evaluate the system performance through a realistic channel model in various scenarios. We show that the proposed method not only greatly increases the spectral efficiency especially in frequency-division duplex (FDD) communication by using partial CSI feedback, but also has near-optimal sum-rate and outperforms other state-of-the-art full-CSI solutions.
\end{abstract}

\begin{IEEEkeywords}
Massive MIMO, Hybrid beamforming, Beam training, Deep learning, Unsupervised learning.
\end{IEEEkeywords}

\IEEEpeerreviewmaketitle
\thispagestyle{empty}

\section{Introduction}

\lettrine[lines=2]{N}{ew} applications such as \gls{IoT} and vehicular communications continuously increase the demands for higher data rate. To face this challenge, massive \gls{MIMO} has become an essential factor in the design of future cellular systems~\cite{what5G}. In massive \gls{MIMO}, the number of antennas in the \gls{BS} scales up to serve several users and it was shown in~\cite{marzetta} that the effects of fast fading and interference vanish when increasing the number of antenna in the \gls{BS}. Higher multiplexing and diversity gain are thus obtained with massive \gls{MIMO}, which in turn results in a higher spectral efficiency and greater energy efficiency.

On the other hand, each antenna in a massive MIMO array requires a \gls{RF} chain. Therefore, the power consumed by \gls{RF} elements such as power amplifiers renders massive \gls{MIMO} systems expensive and energy inefficient. To address this power consumption issue, reduce the cost-power hardware overhead, and yet provide reasonable performance, the \gls{HBF} technique was introduced~\cite{HBF_survey}. It consists of using a small number of analog beamformers deployed to drive  multiple antennas to form a beam, each connected through a single RF chain to a digital precoder. This hybrid combination of phase-based analog and baseband digital beamformers reduce the number of transmission chains while keeping the sum-rate to an acceptable level~\cite{Ayach2014, sohrabi_HBF}. 
In fact, hybrid beamforming techniques have been considered for fifth generation cellular network technology (5G) in the \gls{mm-Wave} bands~\cite{Han2015}. However, an explicit estimate of the \gls{mm-Wave} channel is generally needed to design the hybrid beamforming matrices at the transmitter.  Although several channel estimation techniques for hybrid beamforming have been proposed in the last few years~\cite{cheHB}, channel estimation remains a complicated task due to the hybrid structure of the precoding and to the imperfections of the \gls{RF} chain. Among prior research, the authors in~\cite{HBF_OFDM1} designed hybrid beamforming by considering \gls{OFDM}-based frequency-selective structures. The wideband \gls{mm-Wave} massive MIMO systems was investigated in~\cite{widebandchHBF} to design the hybrid beamforming. In~\cite{unifiedHBF}, the authors designed an analog beamformer based on the second-order spatial channel covariance matrix of a wideband channel. Authors in~\cite{low_comp_HBF} assumed to have perfect knowledge of the \gls{CSI} and proposed a low-complexity hybrid beamforming.  All of the aforementioned methods strongly depend on full knowledge of the \gls{CSI} or channel estimation by using pilots, which increases the signaling overhead and, therefore, reduces the spectral and  energy efficiency of the system especially in \gls{FDD} communication where \gls{CSI} acquisition and feedback is a challenging task~\cite{Bjornson2016}. Therefore, in this paper we propose a system that, instead of considering the full knowledge of \gls{CSI} or channel reciprocity, uses the \gls{RSSI} to design the hybrid beamforming precoders. Unlike \gls{CSI}, \gls{RSSI} is a single real value which users readily measure from the received signal. By consequence, no explicit \gls{CSI} feedback is required, which reduces the signaling overhead and increases the spectral efficiency of the system.

Meanwhile, \gls{DL} techniques have recently been applied widely to telecommunication system~\cite{DNN_WC_Survey} and it was demonstrated to be an outstanding tool for dealing with complex non-convex optimization problems, thanks to its excellent classification and regression capabilities. Although the training process can be time consuming, the training is done in ``off-line''  mode. Therefore, \gls{DL} techniques are a promising approach to reduce the latency in cellular network. Several works have investigated the use of \gls{DNN} to deal with difficult problems within the physical layer~\cite{dl_pl}, including channel coding, channel estimation~\cite{DL-Channel-estimation1}, detection~\cite{DL_detection2} and, beamforming~\cite{Framework, alkh1, HBF, alkh2, HBF_DNNApp}. In particular, the authors of \cite{Framework} considered a multi-input single-output system and solved three optimization problems. By using \gls{DNN} with \gls{FL} and decomposition of the channel matrix in \cite{HBF}, the near-optimal analog and digital precoders have been designed. In~\cite{alkh1,  HBF}, the authors propose a deep supervised learning based method to estimate the hybrid beamforming by knowing the full \gls{CSI}. In~\cite{alkh2}, a supervised deep learning-based hybrid beamforming design for coordinated beamforming is proposed. The authors in \cite{HBF_DNNApp} deployed a \gls{CNN} to design the hybrid beamforming, by knowing the \gls{CSI}. In all the mentioned literature, perfect knowledge of \gls{CSI} is assumed to be available at the \gls{BS}, and the \glspl{DNN} are trained using supervised learning. However, supervised learning requires the optimum targets to be known, and thus requires significant additional computing resources to find these targets using conventional optimization methods. In addition, in practical situations, the knowledge of the optimum  hybrid beamforming structure is hard to obtain. 

Therefore, we introduce in this work a novel approach with \gls{RSSI}-based unsupervised learning to design the \gls{HBF} in massive MIMO system, improved from our first proposal where supervised deep learning was considered~\cite{hojatian2020rssi}. To the best of the authors knowledge, it is the first time that an unsupervised \gls{DL} system is proposed for designing hybrid beamforming precoders in the context of massive MIMO systems. 
Furthermore, we train the DNN specifically for the area where the \gls{BS} is located, so that the geometrical structure of the channel model can be learned. In this paper, this is done using a ray-tracing model~\cite{deepmimo}. The same approach could also be used to train the DNN using direct measurements of the environment.
The proposed DNN architecture is a multi-tasking \gls{CNN} that generates both the analog and digital parts of the hybrid beamforming, enabling to reduce the computational complexity. 
Furthermore, a novel loss function based on sum-rate is proposed. To train the model, we introduce  methods to generate datasets and codebooks based on the deepMIMO channel model~\cite{deepmimo}. Particularly, the synchronization signals transmitted by the \gls{BS} are optimized in such a way that the \gls{RSSI} measurements carry the maximum information about the \gls{CSI}. Three different channel models have been examined to validate the reliability and robustness of the proposed method in different scenarios. These three scenarios have been chosen to cover different environments, received signal strengths and cell coverage. Moreover, we study the effect of \gls{RSSI} quantization on the \gls{DNN}'s performance. The simulation results show that the sum-rate performance of the proposed \gls{RSSI}-based model outperforms other state-of-the-art full-CSI methods while the spectral efficiency, signaling overhead, training time and flexibility of the system are significantly improved.

The major contributions of the paper can be summarized as follows:
\begin{itemize}
    \item a method to design the codebook for the phase-based \gls{AP}; 
    \item a method to design synchronization burst sequences maximizing the channel information carried by the \gls{RSSI}; 
    \item a procedures to generate the \gls{DNN} datasets;
    \item a low complexity approach for fully digital precoder and hybrid beamforming design; and
    \item two unsupervised deep learning methods to directly design the hybrid beamforming.
\end{itemize}

The rest of the paper is organized as follows. Section~\ref{sec:system_model} describes the system model, including beam training during \gls{IA}, RSSI measurement and quantization. In Section~\ref{sec:dataset}, the channel model and dataset generation for \gls{DNN} training are presented, followed by the near-optimal \gls{HBF} solutions for massive MIMO. The proposed method for \gls{SS} planning in \gls{IA} and codebook design for the analog precoder are described in Section~\ref{sec:ss-codebook}. 
In Section~\ref{sec:proposed_DNN}, the \gls{DNN} architecture and the unsupervised learning method are presented.
Finally, in Section~\ref{simulation} the proposed \gls{HBF} methods are evaluated and compared with existing methods, and conclusions are drawn in Section~\ref{Conclusion}.

\textit{Notation:} Matrices, vectors and scalar quantities are denoted by boldface uppercase, boldface lowercase and normal letters, respectively. The notations $(.)^{\rm{H}}$, $(.)^{\rm T}$ , $(.)^{\dagger}$, $|.|$, $\|.\|$, $(.)^{-1}$, $\Re[.]$ and $\Im[.]$ denote Hermitian transpose, transpose, Moore-Penrose pseudoinverse, absolute value, $\ell^2$-norm, matrix inverse, real part, and imaginary part, respectively. 

\section{System Model} \label{sec:system_model}

The considered system model consists of a massive \gls{MIMO} \gls{BS} in a single-cell system equipped with $N_{\sf{T}}$ antennas and $N_{\sf{RF}}$ \gls{RF} chains serving $N_{\sf{U}}$ single-antenna users, as shown in Fig.~\ref{fig:system_arch}. For both uplink and downlink transmission, \gls{HBF} precoders are employed by the \gls{BS}. We consider a fully connected architecture where each \gls{RF} chain is coupled through 2-bit phase shifters to all antennas at the \gls{BS}. A 3-step scenario similar to the one described in~\cite{hojatian2020rssi} is investigated in the following sub-sections. 

\subsection{Step 1: SS Bursts Transmission}

\begin{figure}
    \centering
    \includegraphics[width=0.7\columnwidth]{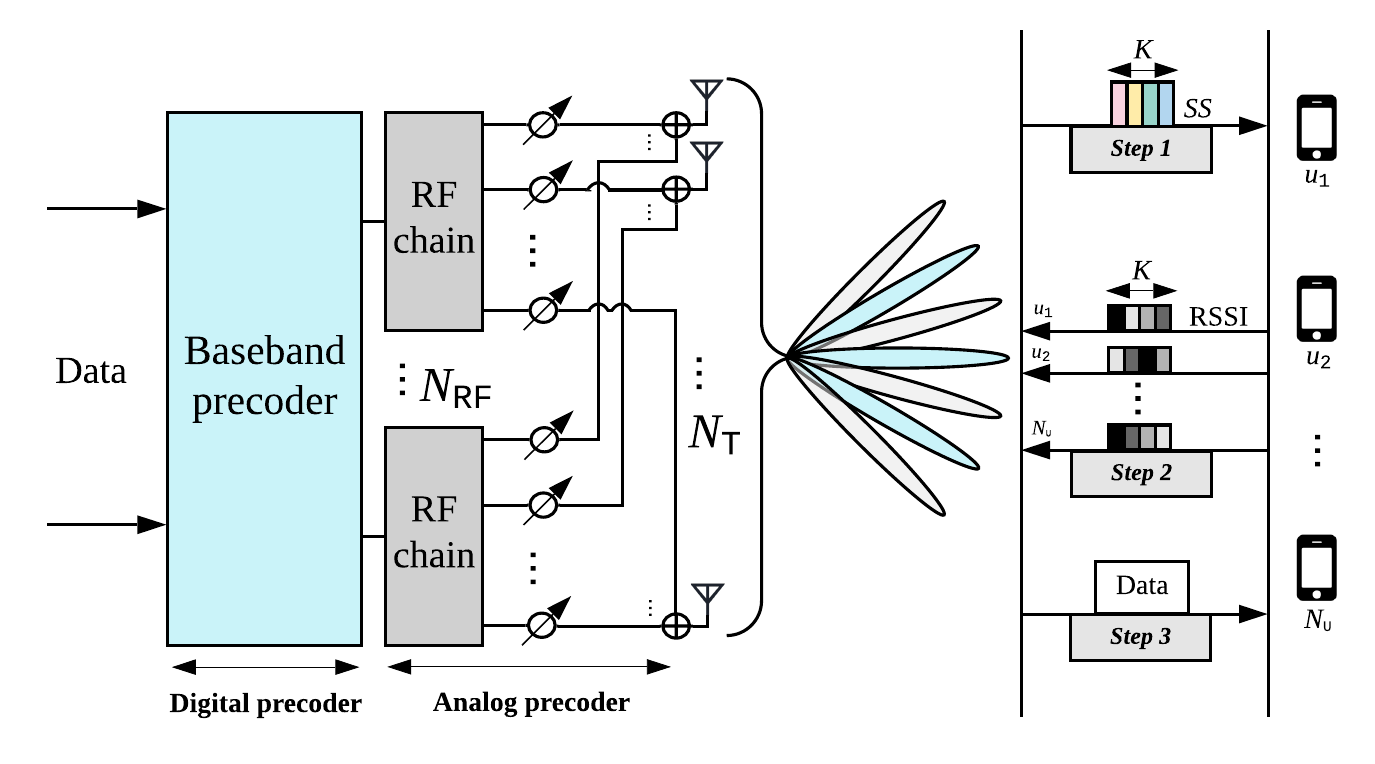}
    \caption{System architecture of the hybrid RF beamforming in \gls{mm-Wave} massive MIMO with three steps beam training}
    \label{fig:system_arch}
\end{figure}

In the first step shown on the right side of Fig.~\ref{fig:system_arch}, the \gls{BS} transmits $K$ \gls{SS} bursts, where each burst $k$ uses different 2-bit phase-shift analog precoders $\mb{A}_{\sf{SS}}^{(k)} \in  \{1,-1,i,-i\}^{N_{\sf{T}} \times 1}$. The \gls{SS} $\mb{A}_{\sf{SS}}^{(k)} s^{(k)}$ are received by all users in the cell. By consequence, the received signal $r_{u}^{(k)}$ at the $u^{\text{th}}$ user for the $k^{\text{th}}$ burst can be written as 
\beq
r_{u}^{(k)} = \mb{h}^{(k){\rm{H}} }_u \mb{A}_{\sf{SS}}^{(k)} + \eta^{(k)}_u,
\eeq
 where $\mb{h}^{(k)}_u \in \mathbb{C}^{N_{\sf{T}} \times 1}$ stands for the channel vector from the $N_{\sf{T}}$ antennas at the BS to user $u$ and $\eta^{(k)}_u$ is the \gls{AWGN} term.
 
 \subsection{Step 2: RSSI Feedback}
 \label{sec:systModel_phase2}
 After receiving $r_{u}^{(k)}$, the averaged \gls{RSSI} value $\alpha_u^{(k)}$ are measured by the $u^{\text{th}}$ user for the $k^{\text{th}}$ \gls{SS} burst, which constitutes the second step. Therefore, we have
\beq \label{alpha}
\alpha_u^{(k)} = \vert  \mb{h}^{(k){\rm{H}} }_u \mb{A}_{\sf{SS}}^{(k)}  \vert^2  + \sigma^2,
\eeq
where $\sigma^2$ is the noise power. All  \gls{RSSI} values of each user are then transmitted to the \gls{BS} through a dedicated error-free feedback channel. These two first steps correspond to the establishment of the \gls{IA} between the \gls{BS} and the users. 

In practical systems, due to the limited precision of the measurements and limitation in feedback channel, the \glspl{RSSI} must be quantized. Let us denote by $\bs{\alpha}_u = [\alpha_u^{(1)}, ..., \alpha_u^{(K)}]^{\rm T}/\beta$ the vector of all \gls{RSSI} values obtained by the $u^{\text{th}}$ user, re-scaled by a factor $\beta$ to ensure that $\alpha_u^{(k)}/\beta \in [0,1] \,\forall k$. Then, we define $\tilde{\bs{\alpha}}_u$ as the quantized \gls{RSSI} vector of user $u$ transmitted to the \gls{BS}. Several quantization methods can be employed. We use linear  quantization, given by
        \beq
        \label{eq:linear_quant}
            \tilde{\bs{\alpha}}_u = \frac{\nint{\bs{\alpha}_u (2^{N_{b}}-1)}}{(2^{N_{b}}-1)},
        \eeq
where $\nint{.}$ is the round operator and $N_{b}$ is the number of quantization bits. We study the effect of quantization on performance in Section~\ref{simulation}.

 \subsection{Step 3: Downlink Data Transmission}\label{sec:systModel_phase3}
The last step corresponds to the downlink transmission where the \gls{BS} transmits data to each user. The digital precoder matrix is $\mb{W} = [\mb{w}_{1}, ..., \mb{w}_{N_{\sf{U}}}]$ where vector $\mb{w}_u \in \mathbb{C}^{N_{\sf{RF}} \times 1}$ is designed to encode the data symbol of user index $u$. The analog precoder $\mb{A} \in  \{1,-1,i,-i\}^{N_{\sf{T}} \times N_{\sf{RF}}}$ is designed to transfer the output of the $N_{\sf{RF}}$ RF chain to $N_{\sf{T}}$ antennas and applies to all users. To reduce the complexity of the \gls{HBF} design, we consider that the analog precoder is chosen from a codebook $\mathcal{A}$ composed of a set of $L$ analog beam codewords $\lbrace \mb{A}_{(1)}, ..., \mb{A}_{(L)} \rbrace$ where $\mathbf{A}_{(l)}$ is the $l^{\text{th}}$ analog precoder matrix of the codebook (the codebook design is discussed  in Section~\ref{sec:ss-codebook}).

The \gls{SINR} of the $u^{\text{th}}$ user received signal for a given hybrid beamformer $(\mb{A},\mb{w}_u)$  is then expressed as 
\begin{align}
    \label{eq:SINR_HBF}
    \text{SINR}(\mb{A},\mb{w}_u) & = \frac{ \big|\mb{h}^{\rm{H}}_{u} \mb{A} \mb{w}_u \big|^2}{\sum_{j \neq u} \big|\mb{h}^{\rm{H}}_{u}\mb{A} \mb{w}_j \big|^2 + \sigma^2},
\end{align}
and the spectral efficiency of the system can be obtained by evaluating the sum-rate expressed as
\begin{align}
\label{eq:sumRate_HBF}
    R(\mb{A},\mb{W}) = \sum_{u=1}^{N_{\sf{U}}}  \text{log}_2 \Bigl(  1+ \text{SINR}(\mb{A},\mb{w}_u) \Bigr).
\end{align}
Then, the \gls{HBF} design consists of finding the digital precoder vectors $\mb{w}_u \; \forall u$ and the analog pecoder matrix $\mb{A}$ in the codebook $\mathcal{A}$ that maximize the sum-rate \eqref{eq:sumRate_HBF}. This problem is however difficult to solve as, in our case, the \gls{BS} does not have a direct knowledge of the channel coefficients $\mb{h}^{\rm{H}}_{u}$ and the noise power $\sigma^2$. The \gls{CSI} is in fact partially embedded in the received \glspl{RSSI}. Therefore, we propose to employ \gls{DNN} techniques to design the  \gls{HBF} precoders. 


\section{Dataset Generation for DNN} \label{sec:dataset}

To train the \glspl{DNN}, a dataset must be obtained beforehand. In practice, this dataset could be generated from channel measurements performed by the \gls{BS}, while in this paper the channel measurements based on the system model described in Section~\ref{sec:system_model} must be simulated. This section describes the channel model and dataset generation procedure followed by the near-optimal full-CSI solution for the \gls{HBF} and \gls{FDP} techniques. The \gls{HBF} and \gls{FDP} described in this section are used as an upper bound to evaluate the unsupervised \gls{DNN} performance.

\subsection{Channel Model}

The deepMIMO~\cite{deepmimo} millimeter-wave massive MIMO dataset model is used to generate the channel coefficients $\mb{h}^{(k)}$ for the train and test datasets. In this model, realistic channel information is generated by applying ray-tracing methods to a three-dimensional model of an urban environment. It provides the channel vector $\mb{h}$ (of length $N_{\sf{T}}$) for each user position on a quantized grid. The considered set of channel parameters from this model are summarized in Table~\ref{tab:deepMIMO_params}. Scenario ``O1'' consists of several users being randomly placed in two streets surrounded by buildings. These two streets are orthogonal and intersect in the middle of the map.

\begin{table}[t]
\centering
\caption{\vspace{1pt}Parameter selection for the deepMIMO channel model}
\vspace{-4pt}
\resizebox{0.4\columnwidth}{!}{
\begin{tabular}{p{1.2cm}c|p{1cm}c}
\toprule
\multicolumn{2}{c}{System}  & \multicolumn{2}{c}{Antennas} \\
Parameter           &  Value & Parameter   &  Value \\
\midrule
scenario  & ``O1''     & num\_ant\_x & $1$      \\
bandwidth & $0.5$ GHz    & num\_ant\_y & $8$    \\
num\_OFDM & $1024$   & num\_ant\_z & $8$      \\
num\_paths& $10$      & ant\_spacing & $0.5$    \\
\bottomrule
\end{tabular}}
\label{tab:deepMIMO_params}
\end{table}

A $N_{\sf{T}} \times N_{\sf{U}}$ channel matrix entry in the dataset is obtained by concatenating  $N_{\sf{U}}$ channel vectors selected randomly from the available user positions of the considered area.

\subsection{Dataset Generation Method}\label{sec:dataSet_gen}

Two datasets are generated for training and testing the network. The first one, referred to as the \emph{core} dataset, contains $N_{core} = 10^4$ channel realizations. This dataset is used to design the codebook $\mathcal{A}$ and the analog precoder of the SS burst $\mb{A}_{\sf{SS}}^{(k)}$. Furthermore, near-optimal  \gls{FDP} and \gls{HBF} precoder solutions are computed to compare the sum-rate performance  of the \gls{DNN}. It is worth mentioning that, as described in Section~\ref{sec:proposed_DNN}, the proposed \gls{DNN} architecture is unsupervised, and does not exploit the near-optimal solutions during training. 

The second dataset contains  $N_\mathrm{DNN} = 10^6$  channel realizations and their related \glspl{RSSI} measured from the $\mb{A}_{\sf{SS}}^{(k)}$ burst generated from the core dataset. It is used to train and test the \gls{DNN}. Note that we have $N_{core} << N_\mathrm{DNN}$ as the core dataset requires resolving computationally heavy optimization problems, and we empirically found that $10^4$ samples is sufficient to obtain good codebook and \gls{SS} design.

\begin{figure}[t!]
    \centering
    \includegraphics[width=0.7\columnwidth]{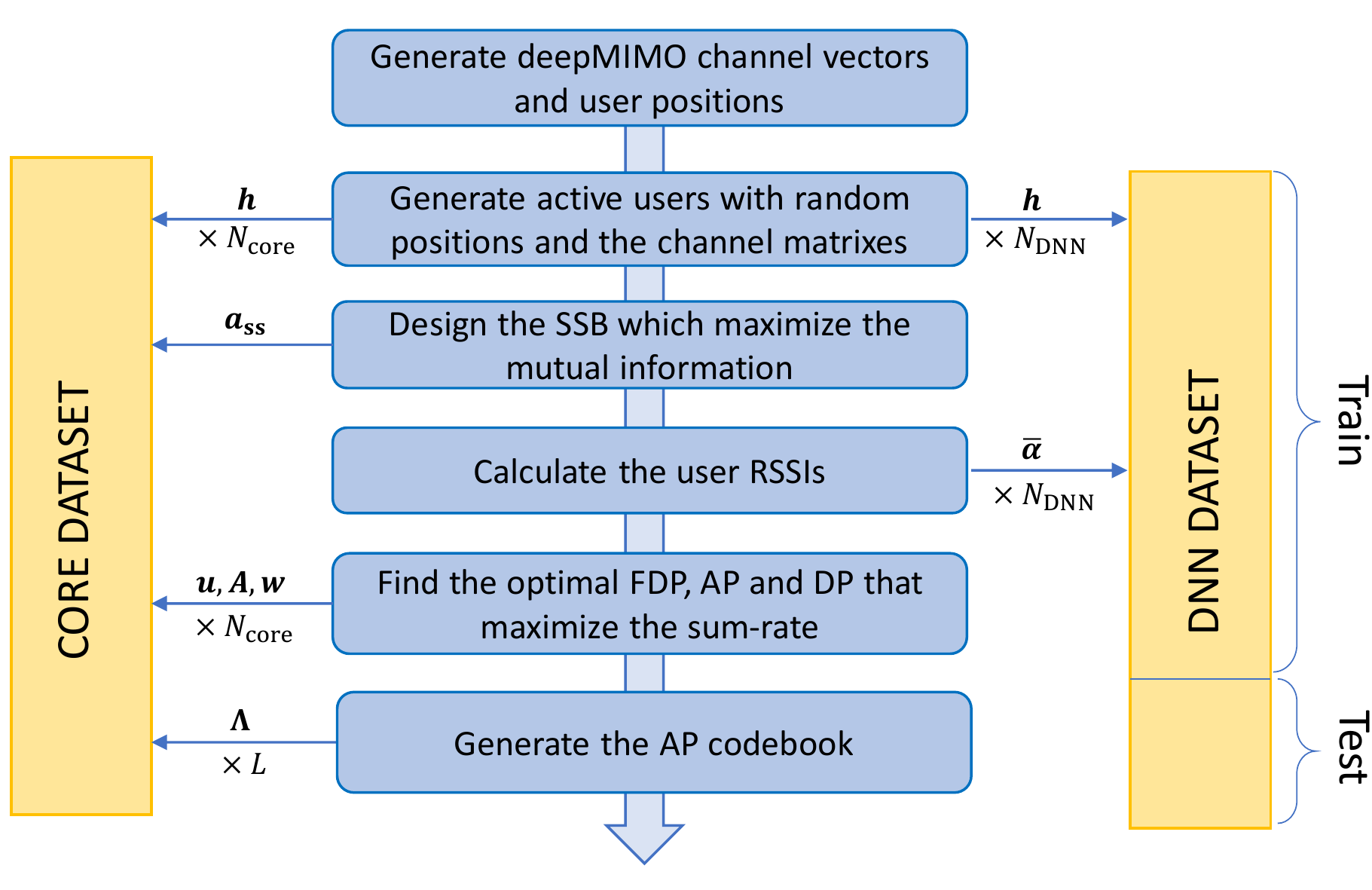}
    \caption{Steps to generate the core and DNN datasets}
    \label{fig:dataset_gen}
\end{figure}

Fig.~\ref{fig:dataset_gen} shows the steps performed to generate both datasets, and the next sections provide more detail about the near-optimal design of the full digital and \gls{HBF} precoders, the codebook generation and the SS bursts design.

\subsection{Fully Digital Precoder Design}
\label{subsec:prop_FDP}

Several \gls{FDP} optimization techniques have been proposed in the literature to maximize the sum-rate. Most of these techniques are computationally heavy when applied for massive \gls{MIMO} cases, but the optimal \gls{FDP} is needed to evaluate the performance of the \gls{DNN}. The \gls{FDP} design problem corresponds to solving the following optimization problem:
\begin{subequations} \label{eq:FDP_opti_problem}
\begin{eqnarray} 
&\underset{\left\lbrace \mb{u}_{u} \right\rbrace}{\max} &
\sum_{u=1}^{N_{\sf{U}}}  \log_2\left(1 + \text{SINR}(\mb{u}_u) \right)  \\
& \text{ s. t.} & \sum_{\forall u} \mb{u}^{\rm{H}} _{u}\mb{u}_{u} \leq P_{\sf{max}}, \label{cnt_op2}
\end{eqnarray}
\end{subequations}
where $P_{\sf{max}}$ stands for the maximum transmission power, and
\begin{align}
    \label{eq:FDP_SINR}
    \text{SINR}(\mb{u}_u) = \frac{ \big|\mb{h}^{\rm{H}}_{u} \mb{u}_u \big|^2}{\sum_{j \neq u} \big|\mb{h}^{\rm{H}}_{u} \mb{u}_j \big|^2 + \sigma^2}.
\end{align}
The method we employed to find the optimal \gls{FDP} is based on~\cite{BBO2014}, where it is demonstrated that the optimal \gls{FDP} vector $\mb{u}_u$ of \gls{FDP} matrix $\mb{U} = [\mb{u}_{1}, ..., \mb{u}_{N_{\sf{U}}}]$ for user  $u$  has the following analytical structure:
%
%
\begin{align}
    \mb{u}_u = \sqrt{p_u} \frac{\Bigl( I_{N_{\sf{U}}} + \frac{1}{\sigma^2}\sum_{i=1}^{N_{\sf{U}-1}}  \mb{h}_i\lambda_i\mb{h}^{\rm{H}}_i \Bigr)^{-1} \mb{h}_u}{ \left\Vert \Bigl( I_{N_{\sf{U}}} + \frac{1}{\sigma^2}\sum_{i=1}^{N_{\sf{U}-1}}  \mb{h}_i\lambda_i\mb{h}^{\rm{H}}_i \Bigr)^{-1} \mb{h}_u \right\Vert},
\end{align}
where $N_{\sf{U}}$ corresponds to the number of users, $I_{N_{\sf{U}}}$ corresponds to the $N_{\sf{U}} \times N_{\sf{U}}$ identity matrix, $p_m$ and $\lambda_m$ are the unknown real-valued coefficients to be optimized, respectively corresponding to the beamforming power and Lagrange multiplier for the user $u$.  In addition, we have $\sum \lambda_u = 1$ and  $\sum p_u = 1$.

Therefore, only $2 \times (N_{\sf{U}}-1)$ real-valued coefficients must be evaluated to resolve the optimization problem, instead of the initial  $N_{\sf{T}} \times N_{\sf{U}}$ complex coefficients. The particle swarm optimization (PSO)~\cite{Trelea2003} algorithm can then be employed to obtain the optimal $\lambda_u$ coefficients. However, we empirically found that near-optimal solutions can be obtained by assuming that $p_u \approx \lambda_u$ and by evenly distributing the power over $\mathcal{K} \in \{1,...,N_{\sf{U}}\}$ users and setting $p_u = 0$ for the remaining $N_{\sf{U}}-\mathcal{K}$ users. Therefore, $2^{N_{\sf{U}}-1}$ solutions have to be evaluated to find the near-optimal one.

\subsection{Hybrid Beamforming Design}
\label{sec:prop_HBF}
When considering hybrid beamforming, \eqref{eq:SINR_HBF} can be rewritten as
\begin{align}
    \text{SINR}(\mb{A},\mb{w}_u) = \frac{ \big|(\mb{A}^{\rm{H}} \mb{h}_{u})^{\rm{H}} \mb{w}_u \big|^2}{\sum_{j \neq u} \big|(\mb{A}^{\rm{H}} \mb{h}_{u})^{\rm{H}}  \mb{w}_m \big|^2 + \sigma^2}.\label{eq:SINR_HBF_eq} 
\end{align}
From \eqref{eq:SINR_HBF_eq}, $\mb{h'} = (\mb{A}^{\rm{H}} \mb{h})^{\rm{H}}$, $\mb{h} = [\mb{h}_{1}, ..., \mb{h}_{N_{\sf{U}}}]$, can be seen as a virtual channel matrix with $N_{\sf{RF}}$ spatial paths. In this virtual system, $\mb{W} = [\mb{w}_{1}, ..., \mb{w}_{N_{\sf{U}}}]$ is the \gls{FDP} matrix of a transmitter with $N_{\sf{RF}}$ virtual antennas. Then, if $\mb{A}$ is known, the optimization method presented in \ref{subsec:prop_FDP} can be re-used to find the optimal digital precoder $\mb{w}$. 
Therefore, the analog precoder $\mb{A}$ must be designed first. We propose to find $\mb{A}$ such that the channel capacity~\cite{T1999} of $\mb{h'}$ is maximized, giving the following integer nonlinear programming problem:

\begin{align}
\label{eq:opti_channel_capacity}
 \max_{\mb{A}}\Biggl(\sum_{i} \max\Bigl( \text{log}_2(\mu \beta_i ),0   \Bigr)\Biggr),
\end{align}
where $\beta_i$ is the $i^{\text{th}}$ eigenvalue of $\mb{h}\mb{h}^{\rm{H}}$, and $\mu$ is the waterfill level chosen to satisfy the following equation:
\begin{align}
 \rho = \sum_{i} \max \Bigl( \mu - \frac{1}{\beta_i} ,0   \Bigr),
\end{align}
with $\rho$ corresponding to the \gls{SNR}. To solve this problem, we used the genetic algorithm~\cite{WD1994}. This iterative algorithm consists, for each iteration (or ``generation''), of selecting, mutating and merging solutions from a set of candidate solutions (referred as ``population'') who gives the best score with respect to the objective function (the sum-rate in our case). At the first generation, $N_c$ candidates are generated randomly. The process of selection consists of keeping, for the next iteration, the $\Psi < N_c $ solutions, referred as elites, which maximize the score. The mutation process makes small random changes on the population, excluding the elites. Finally, the merging process, called crossover, generate $\zeta N_c$ new solutions by mixing solutions from the previous population, with $\zeta$ being the crossover factor. To optimize the analog precoder, we set the genetic algorithm parameters to $N_c = 100 \times N_{\sf{T}}N_{\sf{RF}}$, $\Psi = 0.05  \times N_c$ and $\zeta = 0.8$. This hybrid beamforming design will be referred to as the \gls{HSHO} method.

It is worth noting that the design of the $\mb{A}_{\sf{SS}}^{(k)}$ precoders (step~$1$) has an impact on the amount of channel information carried by the \glspl{RSSI} (step~$2$), which in turn affects the quality of the \gls{HBF} solutions obtained by the \gls{DNN} and the performance of uplink transmission in step~$3$. In addition, the codebook design can also greatly affect the performance of the \gls{DNN}. Therefore, the codebook and SS for initial access need to be carefully planned.

\section{Synchronization Signal and Codebook design} \label{sec:ss-codebook}
In this section, we address the problem of the \gls{SS} burst and analog beamformer codebook design with novel methods.

\subsection{Proposed SS Burst Design} \label{SSB_IA}
\label{sec:prop_SSB}
 As mentioned in Section~\ref{sec:systModel_phase3}, 
the design of the \gls{SS} burst can impact the \gls{HBF} sum-rate performance because it affects the amount of information that is revealed by the \gls{RSSI} measurements about the \gls{CSI}. The \gls{SSB} length also impacts the system data rate in both downlink and uplink. Therefore, it is important to design the \gls{SSB} so that the measured \glspl{RSSI} provide the maximum information about the CSI while minimizing the amount of data to transmit in the feedback channel. 

To resolve this problem, we propose to find the \gls{SS} sequences that maximize the mutual information $I$ between the channel matrices $\mb{h}$ and the quantized \glspl{RSSI} matrix $\tilde{\bs{\alpha}} = [\tilde{\bs{\alpha}}_1,...,\tilde{\bs{\alpha}}_K]$ with $K$ being the number of SS burst
\begin{align}
\label{eq:MI_optim}
I( \mathcal{Q}(\mb{h}), \tilde{\bs{\alpha}} ) = \textsf{H}(\mathcal{Q}(\mb{h})) + \textsf{H}(\tilde{\bs{\alpha}}) - \textsf{H}(\mathcal{Q}(\mb{h}),\tilde{\bs{\alpha}}),
\end{align}
where $\mathcal{Q}(.)$ is the quantization function defined in \eqref{eq:linear_quant}, and where $\textsf{H} (x)$ corresponds to the entropy of the variable $x$ and $\textsf{H}(x_1,...,x_{L})$ is the joint entropy of the variables $\sf{x_l}$, quantized on $Q_l$ values, defined as
\begin{multline}
\textsf{H}( x_1,...,x_L ) = - \sum_{k=0}^{Q_1-1} ... \sum_{z=0}^{Q_L-1} \Bigl( \Pr(x_1 = k, ... , x_L = z) \times \text{log}_2 \; \Pr(x_1 = k, ... , x_L = z) \Bigr),
\end{multline}
with $\Pr$ denoting probability. Furthermore, we have $\textsf{H}(\tilde{\bs{\alpha}}) = \textsf{H}(\tilde{\bs{\alpha}}_1,...,\tilde{\bs{\alpha}}_K)$. 
It is worth mentioning that there is no need to include more than one user in the calculation of  $I$. In fact, user positions are selected randomly and independently, therefore knowing the \gls{RSSI} of one user cannot provide any information for a second user. 

A straightforward computation of \eqref{eq:MI_optim}  can be computationally heavy, particularly when used in an optimization loop and in the case of massive \gls{MIMO}. To reduce the complexity, we assumed that there exists a bijective function that links the channel matrix and the $(X_{\sf{U}},Y_{\sf{U}})$ quantized user positions in the environment. In fact, this assumption is quite realistic due to the high channel diversity induced by massive MIMO systems. Under this assumption, we have $I( \mathcal{Q}(\mb{h}), \tilde{\bs{\alpha}} ) = I( (X_{\sf{U}},Y_{\sf{U}}), \tilde{\bs{\alpha}} ) $. Finally, the genetic algorithm with the same parameters as described in Section~\ref{sec:prop_HBF} is used to design the SS burst sequences $\mb{A}_{\sf{SS}}^{(k)}$, by solving the following optimization problem:
\begin{align}
\max_{\mb{A}_{\sf{SS}}^{(k)} \; \forall \; K} I( (X_{\sf{U}},Y_{\sf{U}}), \tilde{\bs{\alpha}} ).
\end{align}

Note that $\textsf{H}(X_{\sf{U}},Y_{\sf{U}}) \geq \sf{I}( (X_{\sf{U}},Y_{\sf{U}}), \tilde{\bs{\alpha}} ) $ gives the theoretical minimum number of bits required by the feedback channel to transmit all the information about the CSI. Since the user positions are selected randomly with uniform probability on a set of $P_{\sf{U}}$ different locations, we also have $\textsf{H}(X_{\sf{U}},Y_{\sf{U}}) = \text{log}_2(P_{\sf{U}})$. 

\subsection{Proposed Codebook Design}\label{codebook_design}

To reduce the complexity of the \gls{AP} design task for the DNN described in the next section, we propose to restrict the $4^{N_{\sf{T}}N_{\sf{RF}}}$  possible \gls{AP} solutions to a subset (codebook) of $N_{\sf{CB}}$ solutions (codewords), directly chosen during the optimization phases when generating the core dataset. This is achieved through three successive steps described below.

The first step consists of generating a first codebook $\mathcal{A}$ when the near-optimal analog and digital precoder solutions presented in Section~\ref{sec:prop_HBF} are iteratively evaluated for each  channel realization. Let us respectively denote $\mb{A}(n)$ and $\mb{W}(n)$ the \gls{AP} and \gls{DP} solutions related to the $n^{\text{th}}$ channel matrix of the core dataset. Each of these solutions has sum-rate  $R(\mb{A}(n),\mb{W}(n))$, which can be calculated from \eqref{eq:sumRate_HBF}. As a reminder, $\mb{A}_{(l)}$ denotes the $l^{\text{th}}$ analog precoder in the codebook. Then, $\mb{A}(n)$ is appended in the codebook if 
\begin{align}
 R(\mb{A}(n),\mb{W}(n)) > \xi \; \max_{\forall l} \; R(\mb{A}_{(l)},\mb{W}_{(l)}(n)),
\end{align}
where $\xi > 1$ is an arbitrary threshold value, set to $1.005$ in our setup, to decide if the analog precoder will be appended to the codebook, and $\mb{W}_{(l)}(n)$ is the \gls{DP} solution for the $n^{\text{th}}$  channel matrix in the dataset when the analog precoder $\mb{A}_{(l)}$ is chosen. Therefore, each analog precoder solution solved using the genetic algorithm is appended in the codebook if the obtained sum-rate is higher than $\xi$ times the best sum-rate obtained with the \glspl{AP} in the current codebook. Otherwise, the obtained \gls{HBF} solution is replaced by the one in the codebook having the highest sum-rate. To avoid high computational overhead, no \glspl{AP} are appended to the codebook when its size reaches $1000$ \glspl{AP}.

Since the \glspl{AP} are iteratively appended in the first step, it may be possible that better \gls{AP} solutions are found in the later iterations. Therefore, the sum-rate could be improved for the first channel matrices in the dataset. To address this issue, the second step aims to update the \gls{AP} solutions for each channel matrix in the dataset by choosing the \gls{AP} in the codebook, obtained in the first step, that maximize the sum-rate: 
\begin{align}
 \mb{A}{(n)} = \argmax_{\forall \mb{A}_{(l)}\in \mathcal{A}} R(\mb{A}_{(l)},\mb{W}_{(l)}(n))
\end{align}

The final step consists of reducing the size of the codebook while keeping an acceptable level of average sum-rate over the whole dataset. We denote $\mathcal{L}_l$ the set of \gls{AP} index $n \in \{1,...,N_{\sf{core}}\}$ in the core dataset that use the \gls{AP} label number $l$ ($\mb{A}_{(l)}$) in the current codebook as solution. To reduce the codebook size $|\mathcal{A}|$, where $|.|$ represent the cardinality operator in this context, the following operations are iteratively executed:
\begin{itemize}
    \item The codebook is sorted in ascending order such that $ | \mathcal{L}_0 | \leq  |  \mathcal{L}_l | \; \forall \; l$,
    \item All the APs in the core dataset using the AP codeword $A_{(0)}$  (indexed in $\mathcal{L}_0$) are moved to other AP codewords giving the best sum-rate: 
    \begin{align} \mb{A}{(n)} = \argmax_{\mb{A}_{(l)}, l \neq n} R(\mb{A}_{(l)},\mb{W}_{(l)}(n)), \forall n \in \mathcal{L}_0 \end{align}
    \item The AP codeword $0$ is then removed from the codebook. 
\end{itemize}
Each time a codeword is removed, the average sum-rate over the core dataset is reduced. To maintain good performance, the codebook size reduction is stopped when the average sum-rate reaches $99.5$\% of the initial sum-rate.

\section{RSSI-Based Hybrid Beamforming with Deep Neural Network }\label{sec:proposed_DNN}

In this section, we propose a novel \gls{RSSI}-based method to design the \gls{HBF} for massive \gls{MIMO} systems.  The \gls{RSSI} measurements by the users was explained in Section~\ref{sec:systModel_phase2}. The \gls{BS} receives the quantized \gls{RSSI},  $\tilde{\bs\alpha}~=~[\tilde{\bs\alpha}_{1}, ..., \tilde{\bs\alpha}_{N_{\sf{U}}}]$ through a dedicated error-free feedback channel. Then, the BS designs the \gls{HBF} matrices used to transmit data to users in the downlink step, as described in Section~\ref{sec:systModel_phase3}. 
In supervised learning, a computationally heavy optimization problem must be solved not only for each sample in the dataset, but also for each \gls{BS} in the cellular network because the dataset generation depends on the location of the \gls{BS}. On the contrary, in unsupervised learning,   the \gls{BS} can be trained directly using data measured from the environment, thus significantly reducing the computational complexity of the training in the network.


\begin{figure*}[t!]
    \centering
    \includegraphics[width=\textwidth]{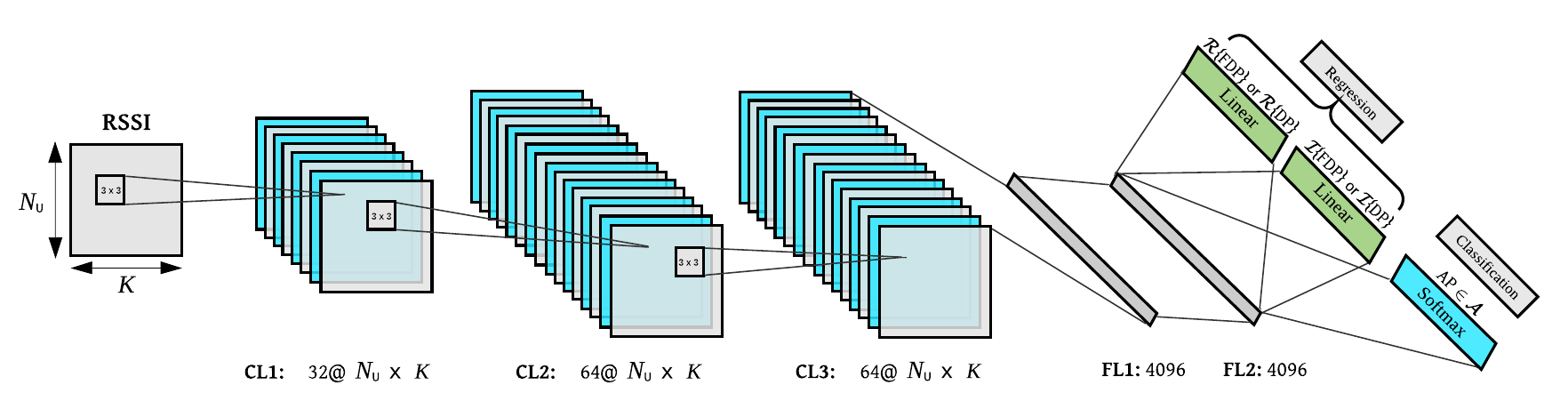}
    \caption{Proposed multi-tasking \gls{DNN} architecture to design the \gls{HBF}}
    \label{fig:CNN_Arch}
\end{figure*}

\subsection{Deep Neural Network Architecture}
Two approaches are considered to obtain the \gls{HBF} solution. In the first approach, we design a DNN, called ``HBF-Net'', to jointly predict the analog precoder $\mb{A}_{(l)} \in \mathcal{A}$ and digital precoder $\mb{W}$. In the second approach, referred to  as ``AFP-Net'', the DNN predicts the analog $\mb{A}_{(l)}$ and fully digital precoders $\tilde{\mb{U}}$. The digital precoder $\tilde{\mb{W}}_{(l)}$ is then computed using
\beq \label{wpre}
\mb{W}_{(l)} =  \mb{A}_{(l)}^{\dagger} \mb{U}.
\eeq 
The AFP-Net approach ensures that the alignment between analog and digital precoders is maintained.
In both approaches, the \gls{DNN} architecture is composed of a classification task for the analog precoder prediction, and of a regression task for the digital precoder prediction. To reduce the number of parameters and accordingly the computational complexity, we design a multi-tasking \gls{DNN}~\cite{multi-task} to perform both tasks in a single \gls{DNN}. The proposed DNN architecture is shown in Fig.~\ref{fig:CNN_Arch}. Referring to Fig.~\ref{fig:CNN_Arch}, the input of the \gls{DL} model is the quantized \glspl{RSSI} received at the BS followed by shared \gls{CL}, \gls{FL} layers and an output layer for each task. It is worth noting that for both HBF-Net and AFP-Net, the \gls{DNN} has the same architecture, except for the dimension of the output layer. Since we separate the real and imaginary parts of the output, the dimension of the output layer for \gls{DP} task in HBF-Net is $2 \times N_{\sf{U}} \times N_{\sf{RF}}$, and $2 \times N_{\sf{U}} \times N_{\sf{T}}$ for \gls{FDP} task in AFP-Net. The dimension of the \gls{AP} task is the same for both approaches and it is equal to size of the codebook ($L = N_{\sf{CB}}$). The \glspl{CL} use  the ``Same'' convolution operators where the dimension of the inputs and outputs are preserved.  After each layer, we use batch normalization to reduce the internal covariate shift and accelerate the learning. In fact, batch normalization is a kind of regularization technique which prevents over-fitting because of its noise injection effect~\cite{batchN2}. Since we have used batch normalization technique, the dropout probability for all layers is selected to a very small value ($0.05$) to avoid the over regularization problem. The dropout layers reduce the impact of the choice of initial weights~\cite{dropout}. 
In addition, we use the leaky ReLU activation function in each layer (except the output layers), to fix the \textit{dying ReLU} problem after batch normalization~\cite{leakyrelu}. This activation function with input $X$ and output $Y$ is given by
\beq
Y =
\begin{cases}
X   &\text{if $X\geq0$}, \\
0.01X &\text{if $X<0$}.
\end{cases}
\eeq
This function does not have a zero-slope part~\cite{leakyrelu}. For the output layer of the classifier or \gls{AP}~task, we use a conventional ``Softmax''  activation function defined as
\beq \label{eq:softmax}
p_{a_{(l)}} = \frac{e^{a_{l}}}{\sum_{j=1}^{L} e^{a_{j}}},
\eeq
where $a_{l}$ is the value of the $l^{\text{th}}$ output of the \gls{DNN} in \gls{AP}~task part and $p_{a_{(l)}}$ is the corresponding output of the Softmax activation function. Thus, the output vector of the classifier after activation function is $\mb{p} = [p_{a_{(1)}}, ..., p_{a_{(l)}}, ..., p_{a_{(L)}}]$ where from Section~\ref{codebook_design} we consider that $L = N_{\sf{CB}}$. Finally,  we used the ``Adam'' algorithm as network optimizer and ``ReduceLROnPlateau'' to schedule the reduction of learning rate~\cite{Adam}. 

As shown in Fig.~\ref{fig:CNN_Arch}, the output of the \gls{FDP} task is separated into real and imaginary parts. Some \gls{DL} framework only support real algebra, and therefore cannot directly support complex-valued computation. Therefore, we separately deploy the real and imaginary part to compute the loss function. Then, to compute the  pseudoinverse of all codewords $\mb{A}_{(l)}$, we use $\mb{A}_{(l)}^{\dagger} = (\mb{A}_{(l)}^{\sf{H}} \mb{A}_{(l)})^{-1} \mb{A}_{(l)}^{\sf{H}}$.
To compute the $(\mb{A}_{(l)}^{\sf{H}} \mb{A}_{(l)})^{-1}$, we define the square matrix $\bs{\Phi} \triangleq \mb{A}_{(l)}^{\sf{H}} \mb{A}_{(l)}$ and  $\bs{\Phi}^{-1} \triangleq \mb{C} + i \mb{D}$ where $\mb{C}$ and $\mb{D}$ are purely real. We then have the following relations based on~\cite{matrix_comp_Inv}:
\begin{align}\label{pseudo}
    \mb{C} &= \left(\Re[\mb{\bs{\Phi}}] + \Im[\mb{\bs{\Phi}}]\Re[\mb{\bs{\Phi}}]^{-1}\Im[\mb{\bs{\Phi}}]\right)^{-1} , \nonumber \\ 
    \mb{D} &= - \Re[\mb{\bs{\Phi}}]^{-1}\Im[\mb{\bs{\Phi}}]\mb{C} \, . 
\end{align}
Eq.~\eqref{pseudo} is valid when the matrix $\Re[\bs{\Phi}]$ is non-singular, which holds in our case. In other cases where $\Re[\mb{\bs{\Phi}}]$ is singular, the corresponding equations are listed in~\cite{matrix_comp_Inv}. To obtain the predicted digital precoders $\mb{W}_{(l)}$ in \eqref{wpre}, we compute
\begin{equation}\label{reim}
\begin{bmatrix}
    \Re[\mb{W}_{(l)}]  & -\Im[\mb{W}_{(l)}]      \\
    \Im[\mb{W}_{(l)}]  & \Re[\mb{W}_{(l)}] 
\end{bmatrix}
= 
\begin{bmatrix}
    \Re[\mb{C}]  &   -\Im[\mb{D}]    \\
     \Im[\mb{D}]  &   \Re[\mb{C}]      
\end{bmatrix}
\begin{bmatrix}
    \Re[\mb{U}]  &   -\Im[\mb{U}]    \\
     \Im[\mb{U}]  &  \Re[\mb{U}]      
\end{bmatrix}
\end{equation}
and then, $\mb{W}_{(l)} = \Re[\mb{W}_{(l)}] + i \Im[\mb{W}_{(l)}]$ is formed.

\subsection{Unsupervised Training}

\begin{figure}[t!]
    \centering
    \includegraphics[width=0.7\columnwidth]{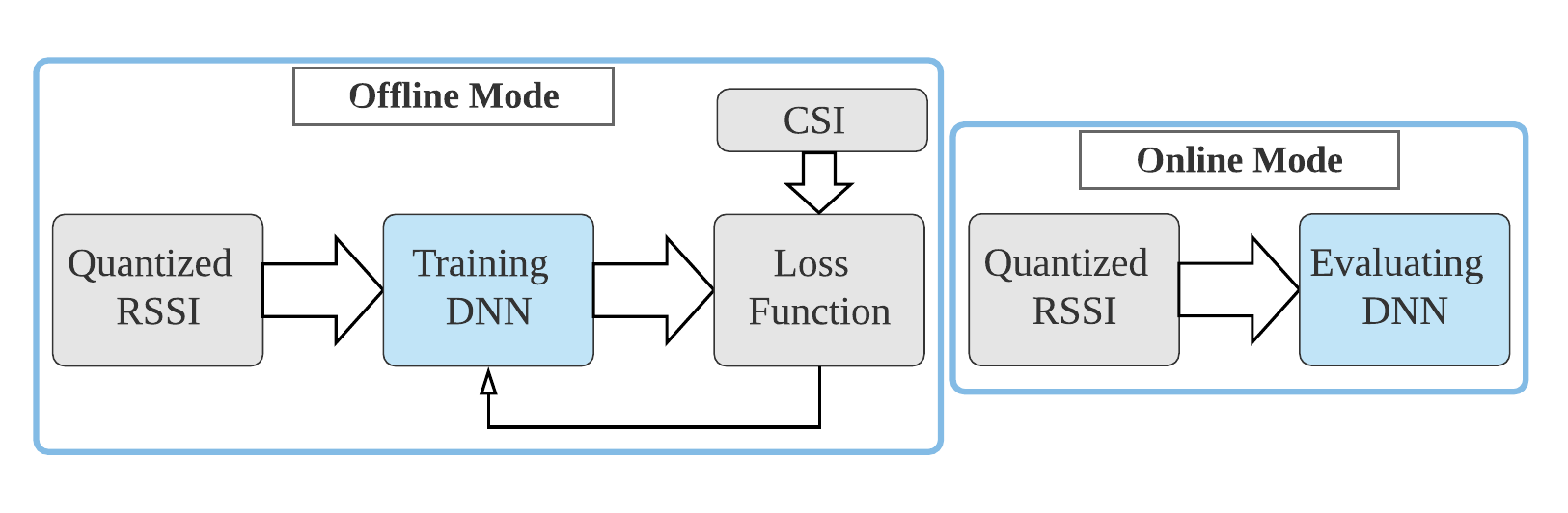}
    \caption{The diagram for the unsupervised training and prediction mode of the proposed \gls{DNN}.}
    \label{fig:unsupervised}
\end{figure}

We propose a novel loss function to train both \gls{AP} task and \gls{DP} or \gls{FDP} tasks without any target. Further, as shown in Fig.~\ref{fig:unsupervised}, the full CSI is only used in training mode to calculate the unsupervised loss function, and in evaluation mode, only the \gls{RSSI} information is used at the \gls{BS} and the \gls{BS} does not have access to \gls{CSI}.  
Since we design the hybrid beamforming with two different approaches, each approach requires its own loss function for training. As a reminder, the sum-rate achieved by the \gls{HBF} is given by \eqref{eq:sumRate_HBF}, and based on \eqref{eq:softmax}, we define $\mb{p} = [p_{a_{(1)}}, ..., p_{a_{(l)}}, ..., p_{a_{(L)}}]$ as the output vector of the \gls{AP} task. We define below the unsupervised loss function for each approach.

\subsubsection{HBF-Net}
The \gls{DNN} in this approach is trained to jointly design the \gls{DP} and \gls{AP} directly from \glspl{RSSI}. For this approach we define the unsupervised loss function as
\beq \label{loss_HBF}
\mathcal{L}_{\text{HBF}} = - \sum_{l=1}^{L} p_{a_{(l)}} R(\mb{A}_{(l)},\Bar{\mb{W}}),
\eeq
where $\mb{A}_{(l)}$ is the $l^{\text{th}}$ analog precoder of the codebook and $\Bar{\mb{W}} = [\Bar{\mb{w}}_{1}, ..., \Bar{\mb{w}}_{N_{\sf{U}}}]$ is the output of the \gls{DP} task. To satisfy $||\mb{A}_{(l)} \Bar{\mb{w}}_{u}||^2 = 1$ we further normalize $\Bar{\mb{w}}_{u}$. The negative sign allows the sum-rate to be maximized when the \gls{DNN} is trained to minimize the loss function. Algorithm~\ref{alg:HBFNet_train} summarizes the steps to train the HBF-Net.

\begin{algorithm}[t]
\DontPrintSemicolon
  
  \KwInput{$\tilde{\bs\alpha}$}
  \KwOutputr{DP: $\Re[\Bar{\mb{W}}]$ and $\Im[\Bar{\mb{W}}]$}
  \KwOutputc{AP: $\mb{p}$}

\For{$i$ \textbf{in range(}epochs)}
{
\For{$l$ \textbf{in range(}L)}{\textbf{Compute} $R(\mb{A}_{(l)},\Bar{\mb{W}})$
}
\textbf{Compute}  $\mathcal{L}_{\text{HBF}}$ as in \eqref{loss_HBF}\\
\textbf{Compute} gradient over layers \\
\textbf{Update} weights and biases with \textbf{Adam} optimizer
}
\label{alg:HBFNet_train}
\caption{Training mode in HBF-Net}
\end{algorithm}

\subsubsection{AFP-Net}
The loss function in this approach is different because here we aim to design the digital precoder from the \gls{FDP} and \gls{AP}. To do so, we first obtain the \gls{FDP} and \gls{AP}, and then by using \eqref{wpre}, \gls{DP} can be computed. If we define
\beq \label{ratefdp}
R(\mb{U}) = \sum_{u=1}^{N_{\sf{U}}}  \log_2\!\left(1 + \text{SINR}(\mb{u}_u) \right),
\eeq
the loss function for the \gls{FDP} task can be defined as
\beq \label{loss_FDP}
\mathcal{L}_{\text{FDP}} = -R(\mb{\Bar{U}}) ,
\eeq
where $\Bar{\mb{U}} = [\Bar{\mb{u}}_{1}, ..., \Bar{\mb{u}}_{N_{\sf{U}}}]$ is the output of \gls{FDP} task in AFP-Net. This loss function results in the maximization of the \gls{FDP} sum-rate. Here again, we normalize $\Bar{\mb{u}}_{u}$ to satisfying  $||\Bar{\mb{u}}_{u}||^2 = 1$. Now, by knowing the \gls{FDP}, given \gls{AP}, we compute $R(\mb{A}_{(l)},\tilde{\mb{W}}_{(l)})$, where $\tilde{\mb{W}}_{(l)}$ is the \gls{DP} matrix obtained from \eqref{wpre} for the $l^{\text{th}}$ codeword in the codebook. The loss function for the \gls{AP} task is defined as  
\beq \label{loss_AP}
\mathcal{L}_{\text{AP}} = - \sum_{l=1}^{L} p_{a_{(l)}} R(\mb{A}_{(l)},\tilde{\mb{W}}_{(l)}),
\eeq
where unlike \eqref{loss_HBF}, this loss function is only tuning the Softmax outputs $p_{a_{(l)}}$ of the \gls{AP} task. The final loss function for AFP-Net is defined as 
\beq\label{loss_Uns}
\mathcal{L}_{\text{AFP}} = \mathcal{L}_{\text{FDP}} + \mathcal{L}_{\text{AP}},
\eeq
so that the \gls{DNN} maximizes the \gls{FDP} sum-rate while also maximizing the \gls{HBF} sum-rate by selecting the appropriate \gls{AP} from the codebook.  Algorithm~\ref{alg:AFPNet_train} summarizes the steps to train the AFP-Net. 

It can be seen that AFP-Net is more complex than HBF-Net, which directly outputs the hybrid beamforming solution. However, as shown in Section~\ref{simulation}, AFP-Net achieves better sum-rate performance by keeping the alignment between the analog and digital precoder.

\begin{algorithm}[t]
\DontPrintSemicolon
  
  \KwInput{$\tilde{\bs\alpha}$}
  \KwOutputr{FDP: $\Re[\Bar{\mb{U}}]$ and $\Im[\Bar{\mb{U}}]$}
  \KwOutputc{AP: $\mb{p}$}

\For{$i$ \textbf{in range(}epochs)}
{
\textbf{Compute} $\mathcal{L}_{\text{FDP}}$ as in (\ref{loss_FDP})

\For{$l$ \textbf{in range(}L)}
{\textbf{Compute} $\tilde{\mb{W}}_{(l)} =  \mb{A}_{(l)}^{\dagger} \Bar{\mb{U}}$ as in (\ref{pseudo}) and (\ref{reim})\\

\textbf{Compute} $R(\mb{A}_{(l)},\tilde{\mb{W}}_{(l)})$
}
\textbf{Compute} $\mathcal{L}_{\text{AP}}$ as in \eqref{loss_AP}\\
Total loss = $\mathcal{L}_{\text{FDP}}$ + $\mathcal{L}_{\text{AP}}$\\
\textbf{Compute} gradient over layers \\
\textbf{Update} weights and biases with \textbf{Adam} optimizer
}
\label{alg:AFPNet_train}
\caption{Training mode in AFP-Net}
\end{algorithm}

\subsection{Evaluation Phase}

In the evaluation phase, a part of the \gls{DNN} dataset is dedicated to test the network, as presented in Section~\ref{sec:dataSet_gen}. We used the sum-rate metric to characterize the performance.  For the evaluation phase,  the maximum value  of the ``Softmax'' output is selected in the classifier. So, we can compute the sum-rate of \gls{HBF} to evaluate its performance using $R(\hat{\mb{A}},\hat{\mb{W}})$, where $\hat{\mb{A}}$ is the analog precoder predicted by the \gls{AP} task, expressed as $\hat{\mb{A}} =  \mb{A}_{(\gamma)},$ where $\mb{A}_{(\gamma)} \in \mathcal{A}$,  $\gamma = \argmax (\mb{p})$ and  $\hat{\mb{W}} = [\hat{\mb{w}}_{1}, ..., \hat{\mb{w}}_{N_{\sf{U}}}]$ is the predicted digital precoder of HBF-Net or AFP-Net obtained from (\ref{wpre}).
To satisfy the power constraint we normalize the $\hat{\mb{w}}_{u}$ to have $||\hat{\mb{A}} \hat{\mb{w}}_{u}||^2 = 1$. 

Likewise, to evaluate the \gls{DNN} performance for \gls{FDP} sum-rate we compute $R(\hat{\mb{U}})$, where $\hat{\mb{U}} = [\hat{\mb{u}}_{1}, ..., \hat{\mb{u}}_{N_{\sf{U}}}]$ is the predicted \gls{FDP} matrix in FDP task of AFP-Net. As discussed in the next section, we evaluate the \gls{DNN} using different scenarios.

\section{Numerical Evaluation} \label{simulation}

\begin{figure}[t!]
    \centering
    \includegraphics[width=0.7\columnwidth]{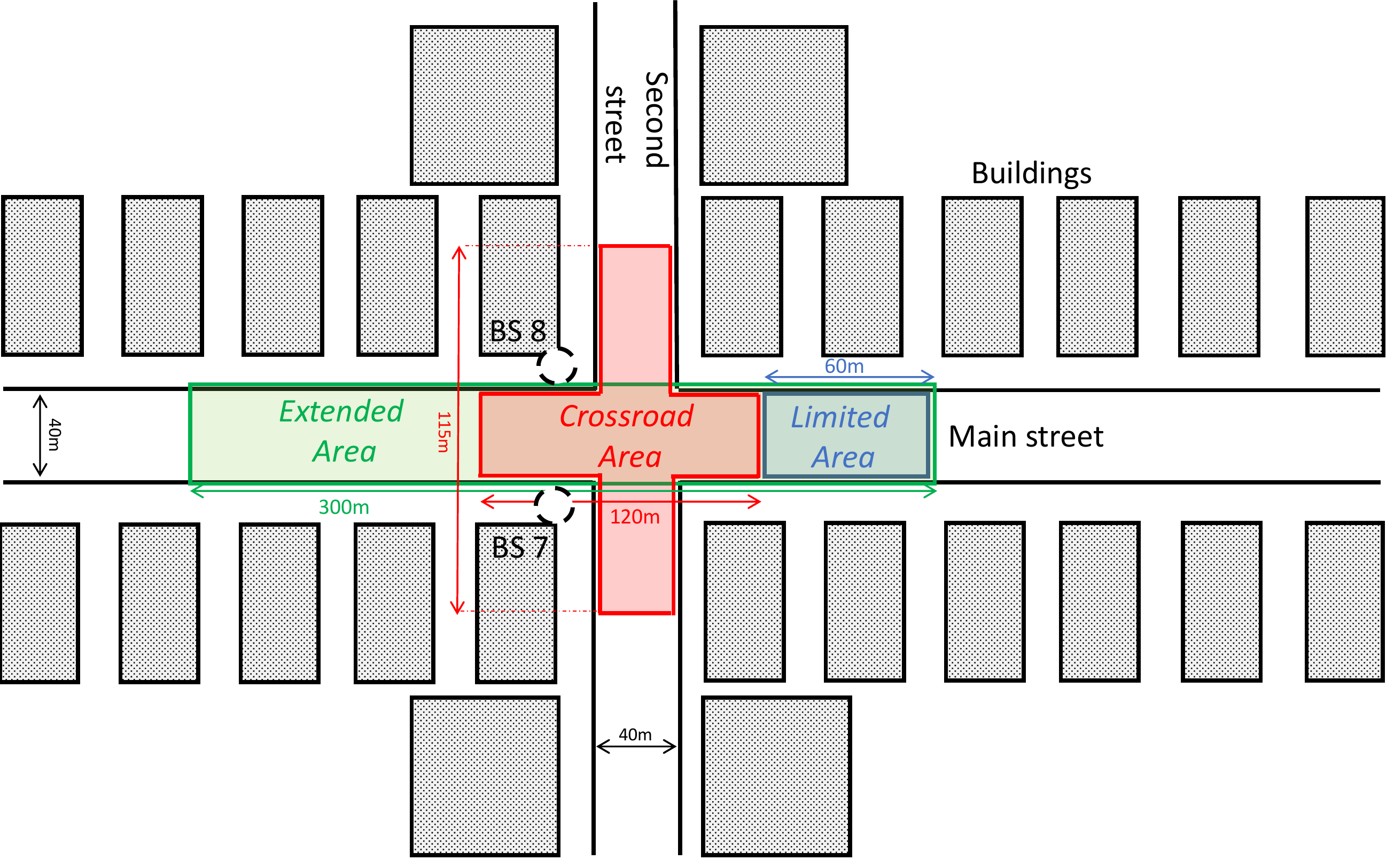}
    \caption{Illustration of each type of area covered for the deepMIMO channel model.}
    \label{fig:deepMIMO_area}
\end{figure}

In this section, the performance of the proposed DNN, implemented using the \textsc{PyTorch} \gls{DL} framework, is numerically evaluated. To analyse how the \gls{DNN} performance  evolves with the environment, three types of datasets covering different areas are considered, as illustrated in Fig.~\ref{fig:deepMIMO_area}: 
\begin{enumerate}
    \item \emph{Limited Area}: small area of $54481$ possible user positions in the main street, with base station number $7$ as the transmitter. The ``active\_user\_first'' ($\text{AUF}$) and ``active\_user\_last'' ($\text{AUL}$) deepMIMO parameters are set to $\text{AUF} = 1000$ and $\text{AUL} = 1300$.
	\item \emph{Extended Area}: larger area than the limited one, with $271681$ possible user positions in the main street, with $\text{AUF} = 1000$, $\text{AUL} = 2500$, and base station number $7$ as the transmitter.
	\item \emph{Crossroad Area}: area located at the intersection of the two streets, with users coming from every direction ($280102$ positions). Three deepMIMO channel environments have been generated and concatenated to obtain this last dataset. The first environment uses $\text{AUF} = 1300$, $\text{AUL} = 1900$, the second uses $\text{AUF} = 3700$, $\text{AUL} = 3852$  and the third uses $\text{AUF} = 3853$, $\text{AUL} = 4300$. For all these models, the base station number $8$ corresponds to the transmitter.
\end{enumerate}
 
The size of the DNN dataset is set to $10^6$ samples for each scenario, with 85\% of the samples used for training as training set and the remaining ones are used to evaluate the performance as test set. Table~\ref{tbl:HP-DNN} shows the chosen hyperparameters which have been used for the proposed multi-tasking \gls{DNN}. 

\begin{table}[t]
    \centering
    \caption{Multi-tasking DNN Hyper-Parameters}
    \resizebox{0.4\columnwidth}{!}{
    \begin{tabular}{lc}
        \toprule
        \multicolumn{1}{c}{Parameter} & \multicolumn{1}{c}{Set Value} \\
        \cmidrule(lr){1-1} \cmidrule(lr){2-2}         
        Mini-batch size& 500 \\
        Initial learning rate &  0.001 \\
        ReduceLROnPlateau (factor) & 0.1 \\
        ReduceLROnPlateau (patience) & 3 \\
        Weight decay & $10^{-6}$ \\ 
        Dropout keep probability & .95 \\
        Kernel size & 3 \\
        Zero padding & 1 \\
        ``$\epsilon$'' in BatchNorm (1D\&2D) & $10^{-5}$\\
        \bottomrule
    \end{tabular}}
    \label{tbl:HP-DNN}
\end{table}

\subsection{Sum-Rate Evaluation for All Areas}

\begin{figure}
    \centering
    \includegraphics[width=0.6\columnwidth]{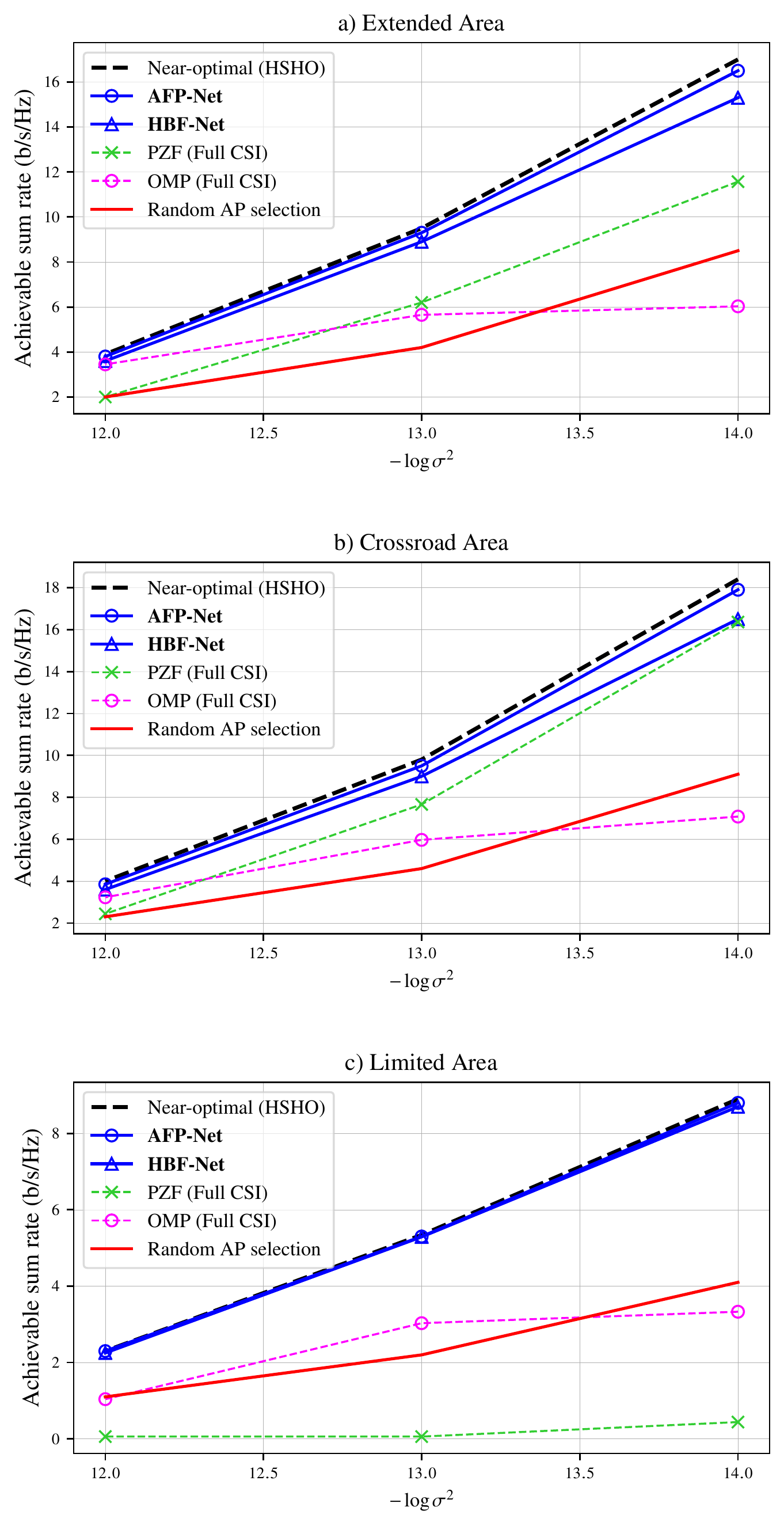}
    \caption{Sum-rate performance of Hybrid beamforming design in ``AFP-Net'' and ``HBF-Net'' in three scenario: a) Extended Area b) Crossroad Area c) Limited Area ($N_{\sf{U}}=4, N_{\sf{RF}}=8, N_{\sf{T}}=64,  K=32$).}
    \label{fig:three_area}
\end{figure}

We consider a system with $4$ users communicating with a BS equipped with $64$ antennas, $8$ RF chains, and $K = 32$ full precision RSSIs are fedback to the BS. Fig.~\ref{fig:three_area} shows the achievable sum-rate for each considered area when considering different noise power values $\sigma^2$ ranging from $-120$~dBW to $-140$~dBW. When considering the channel attenuation, the average \glspl{SNR} for $\sigma^2 \in \{-120, -130, -140\}$~dBW in Extended Area are $10.8$~dB, $20.8$~dB and $30.8$~dB, in Crossroad Area are $10.6$~dB, $20.6$~dB and $30.6$~dB, and in Limited Area are $4.35$~dB, $14.35$~dB and $24.35$~dB, respectively. It can be seen that the AFP-Net has better performance when compared to the HBF-Net. It is owing to the fact that in the AFP-Net, the digital precoder is extracted from the predicted FDP and HBF based on \eqref{wpre}. Therefore, the alignment between analog and digital precoder is preserved. In fact, the sum-rate performance of the AFP-Net is very close to the upper bound obtained by using the full-CSI HBF design method presented in Section~\ref{sec:prop_HBF}.

Furthermore, as shown in Fig.~\ref{fig:three_area} the two proposed unsupervised HBF design methods presented in Section~\ref{sec:proposed_DNN} have better sum-rate performance, for all three areas, when compared with the \gls{PZF}~\cite{low_comp_HBF} and the \gls{OMP}~ \cite{Ayach2014} full CSI techniques. The \gls{OMP} technique uses as inputs the optimal fully digital precoder and the same AP codebook used by the \gls{DNN} and designed according to the method proposed in Section~\ref{sec:ss-codebook}. In fact, the \gls{PZF} has very poor performance for the Limited Area, which is located far from BS, and therefore suffers at low \gls{SNR} level. Unlike \gls{PZF} which has  good performance in high SNR, \gls{OMP} achieves better sum-rate in low \gls{SNR}. However, our proposed near-optimal \gls{HSHO} solution and unsupervised learning \gls{DNN} methods have stable performance in all SNRs. For instance, with $\sigma^2=-130$~dBW, our proposed method in AFP-Net is better than \gls{PZF} by 50\%, 25\% and 8733\% for scenario (a), (b) and (c), respectively, and better than \gls{OMP} by 66\%, 59\% and 75\%.  It is worth mentioning that both \gls{PZF} and \gls{OMP} techniques require perfect knowledge of the \gls{CSI}, and therefore require a high bandwidth feedback channel to report the full \gls{CSI} in \gls{FDD} scenario, which penalizes the spectral efficiency of the system.

The ``Random AP selection'' curves in Fig.~\ref{fig:three_area} show the achieved sum-rate when the BS selects the \gls{AP} from the codebook randomly, while obtaining the \gls{DP} by substituting into \eqref{wpre} the random \gls{AP} and the \gls{FDP} predicted by AFP-Net. The significant performance degradation observed for the Random AP selection confirms that the \gls{AP} task of the \gls{DNN} actually learns the system characteristics from the dataset. 

\begin{figure}
    \centering
    \includegraphics[width=0.6\columnwidth]{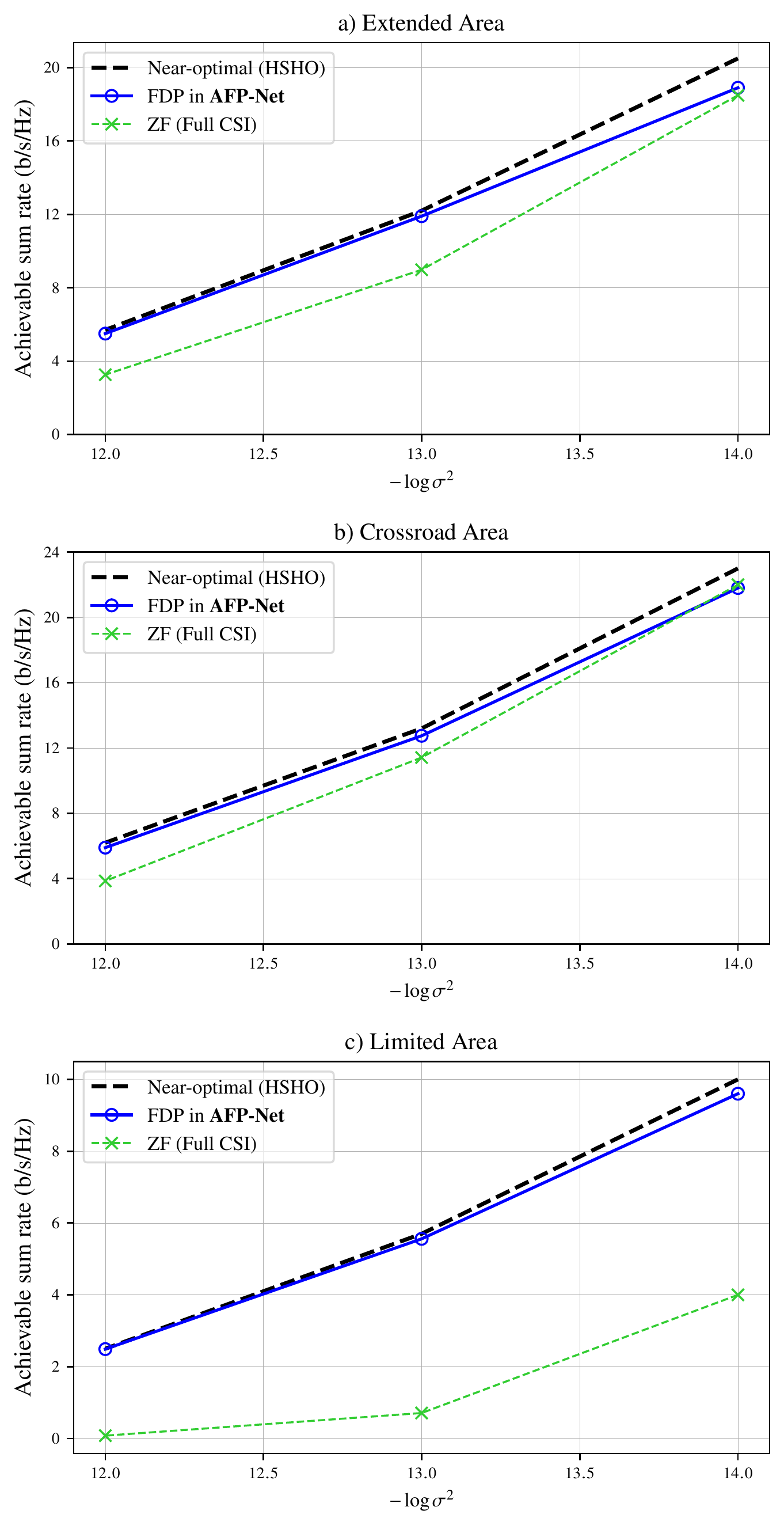}
    \caption{Sum-rate performance of FDP design in ``AFP-Net'' in three scenario: a) Extended Area b) Crossroad Area c) Limited Area ($N_{\sf{U}}=4, N_{\sf{RF}}=8, N_{\sf{T}}=64,  K=32$).}
    \label{fig:three_area_FDP}
\end{figure}

Likewise, Fig.~\ref{fig:three_area_FDP} shows that the sum-rate performance of the FDP obtained with the proposed AFP-Net method outperforms the \gls{ZF} method, for all areas and different noise power. In fact, the proposed unsupervised learning in AFP-Net has near-optimal FDP sum-rate performance in all selected areas. This make the proposed network a promising solution for FDP design in mm-wave massive MIMO systems.

\subsection{Impact of The RSSI Length and Quantization}

\begin{figure}
    \centering
    \includegraphics[width=0.6\columnwidth]{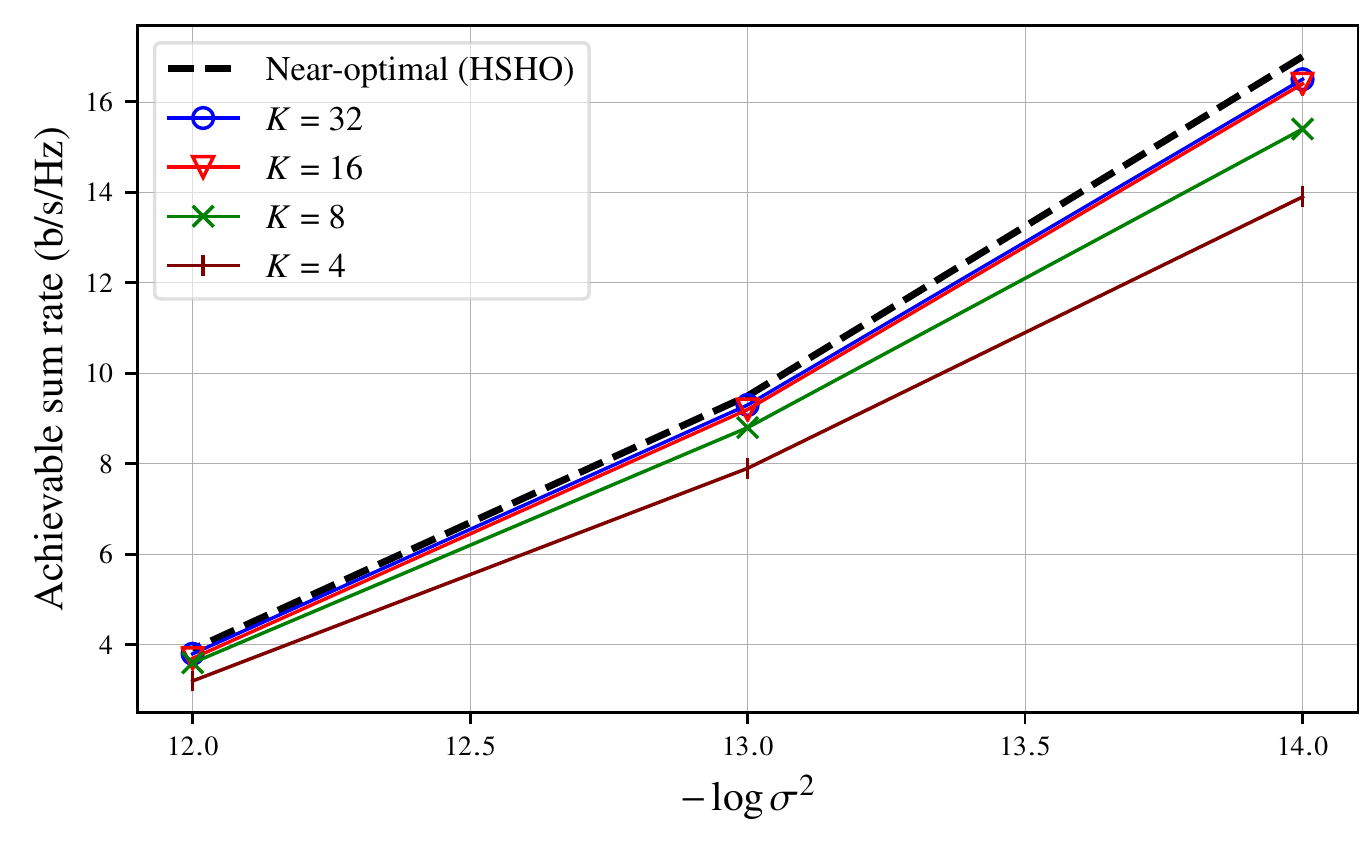}
    \caption{Sum-rate performance of HBF design in AFP-Net in different number of RSSI ($K$) ($N_{\sf{U}}=4, N_{\sf{RF}}=8, N_{\sf{T}}=64$).}
    \label{fig:K_test}
\end{figure}

The effect of the number $K$ of \glspl{SSB} (also the number of RSSI values) on AFP-Net is shown in Fig.~\ref{fig:K_test}, for the ``Extended Area'' scenario. 
Increasing $K$ increases the amount of CSI available to the \gls{DNN}, and we see that the achievable rate indeed improves as $K$ increases. However, increasing $K$ also reduces the channel efficiency because of the need to send more \glspl{SSB} in Step~$1$. Furthermore we see that the performance of AFP-Net with $K = 8$ is close to the performance obtained with $K=16$ and $K=32$.

\begin{figure}
    \centering
    \includegraphics[width=0.6\columnwidth]{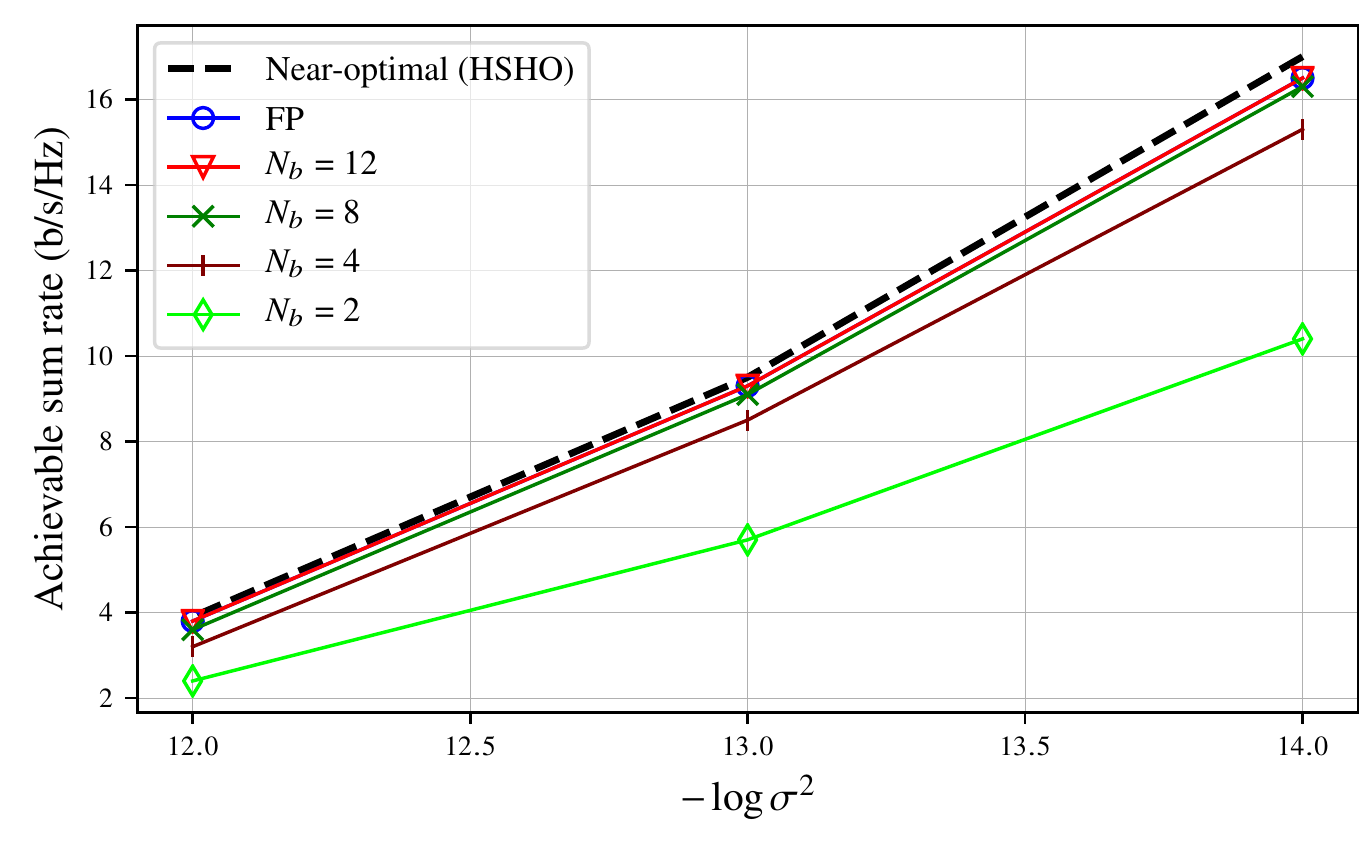}
    \caption{Sum-rate performance of HBF design in ``AFP-Net'' in different number of quantization bits ($N_b$) ($N_{\sf{U}}=4, N_{\sf{RF}}=8, N_{\sf{T}}=64,  K=32$).}
    \label{fig:Nb_test}
\end{figure}

In Fig.~\ref{fig:Nb_test} we examined the proposed methods with different numbers of quantization bits $N_b$ for the RSSIs, computed using \eqref{eq:linear_quant}.  The ``Extended Area''  is considered to evaluate the performance. In fact, there is a trade-off between the spectral efficiency improvement of the system and the performance of the DNN when changing the number of bits. It can be seen in Fig.~\ref{fig:Nb_test} that the sum-rate performance of the DNN for $N_b = 12$ to $N_b = 8$ is very close to the DNN performance when considering full precision for the RSSIs.

\subsection{Impact of The Number of Users and Antennas}

\begin{figure}
    \centering
    \includegraphics[width=0.6\columnwidth]{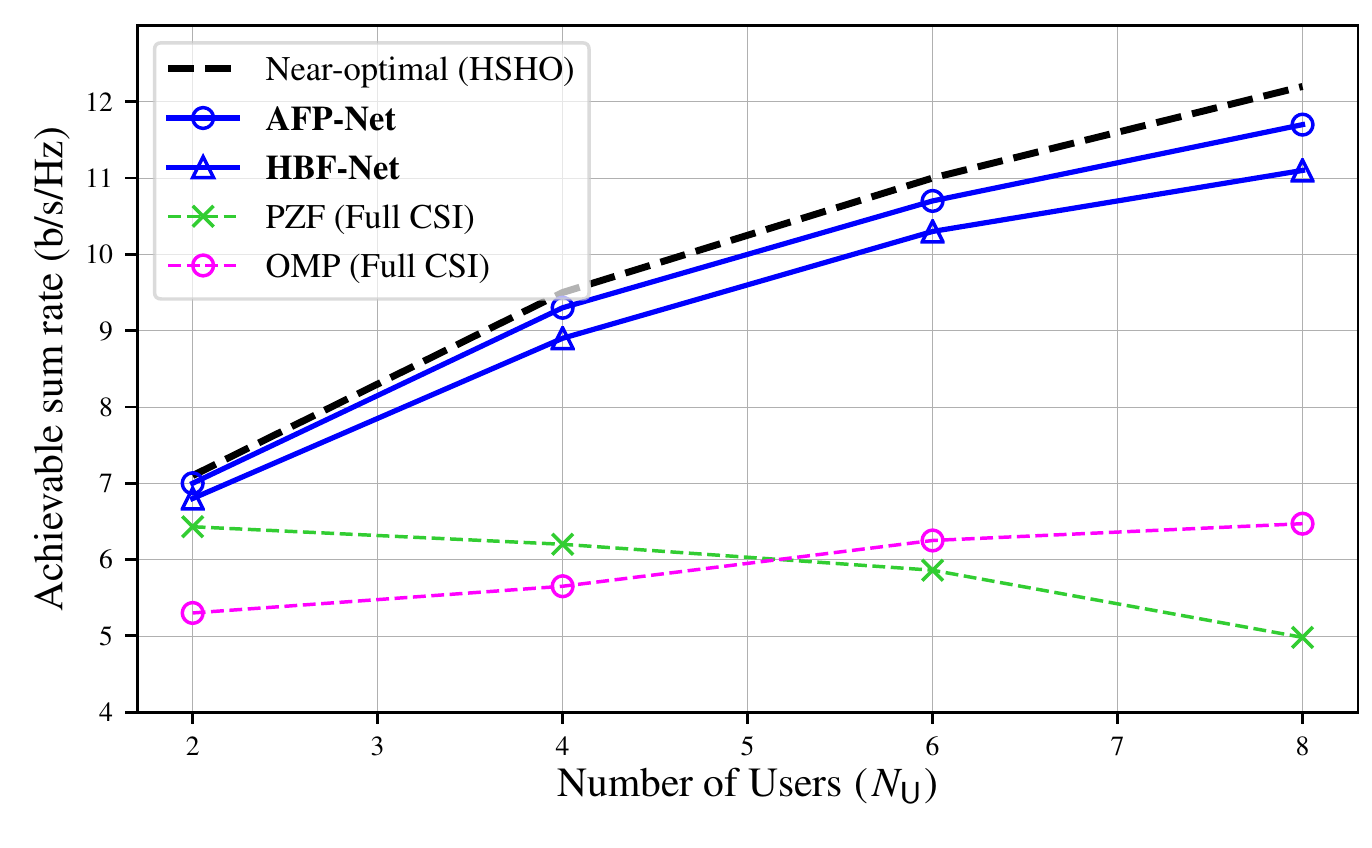}
    \caption{Sum-rate performance of HBF design in ``AFP-Net'' and ''HBF-Net'' versus different number of user ($N_{\sf{U}}$) in ``Extended area'' ($\sigma^2= -130$~dBW, $K=32, N_{\sf{T}}=64$).}
    \label{fig:NU_test}
\end{figure}

\begin{figure}
    \centering
    \includegraphics[width=0.6\columnwidth]{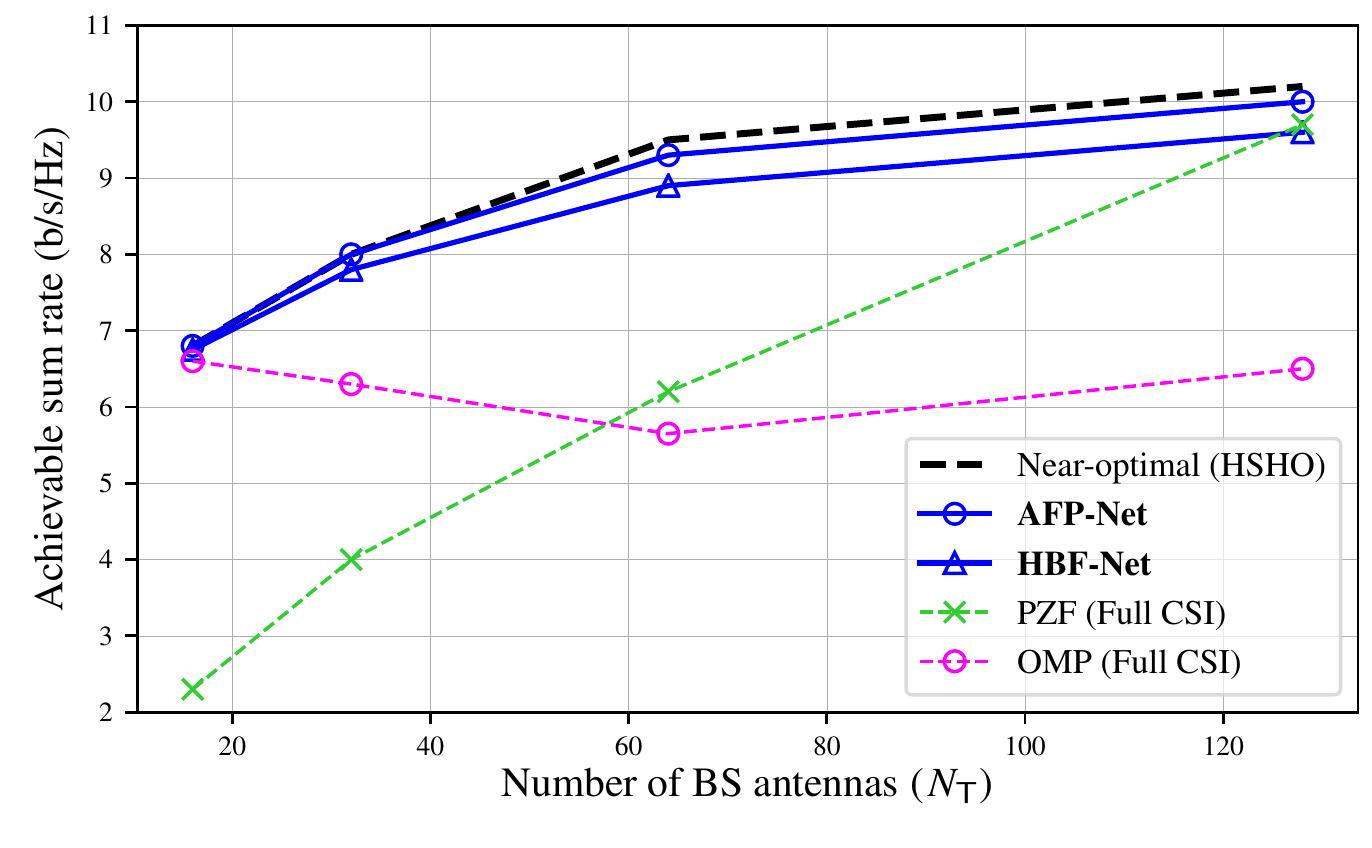}
    \caption{Sum-rate performance of HBF design in ``AFP-Net'' and ''HBF-Net'' versus different number of antenna in BS ($N_{\sf{T}}$) in ``Extended area'' ($\sigma^2= -130$~dBW, $K=32, N_{\sf{U}}=4$).}
    \label{fig:NT_test}
\end{figure}

This section evaluates how the performance of the two proposed HBF designs scale when varying the number of users   $N_{\sf{U}} \in \{2, 4, 6, 8\}$ and the number of antennas $N_{\sf{T}} \in \{16, 32, 64, 128\}$. The number of RSSIs is set to $K=32$, the noise power is fixed to $-130$ dBW and the ``Extended Area''  is considered. In fact, the complexity of the \gls{DNN}, measured in terms of the number of parameters, can be expected to depend on the complexity of the optimization problem. Thus increasing the number of antennas in \gls{BS} or the number of users should require a more complex DNN. However, the results in Figures~\ref{fig:NU_test} and \ref{fig:NT_test} show that the proposed architecture is complex enough to have excellent performance for a wide variety of number of antennas and number of users. Fig.~\ref{fig:NU_test} shows that although our method is RSSI based, it has better performance in comparison with other CSI-based methods and the sum-rate performance scale with the number of users. For instance, our proposed method is better than \gls{PZF} by 50\%, 84\% and 122\% for $N_{\sf{U}} = 4, 6, 8$, respectively. Meanwhile, the \gls{PZF} and \gls{OMP} method do not scale well when the number of users increases. This can be explained by the fact that increasing the number of user and fixing the number of antennas in \gls{BS} will generate more inter-user interference. Furthermore, \gls{PZF} only provides near-optimal results if the ratio between the number of antennas and the number of users is large enough in massive MIMO systems. By increasing the number of users, this ratio decreases and the sum-rate performance of \gls{PZF} collapses.

In Fig.~\ref{fig:NT_test}, we evaluate our proposed method with different number of antennas in BS ($N_{\sf{T}}$). Again, we consider the ``Extended Area'' scenario where $\sigma^2= -130$~dBW, $K=32$ and  $N_{\sf{T}}=64$. As we can see, the \gls{PZF} method has poor performance when the number of antennas is low whereas the sum-rate performance is near-optimal when the number of antennas grows up to $256$. Therefore, our proposed methods outperforms \gls{PZF} for the $16$, $32$ and $64$ antennas cases by 195\%, 100\% and 50\%, respectively. Unlike \gls{OMP} which has fair enough performance with a small number of antennas, the sum-rate of \gls{PZF} is almost the same as our proposal for higher number of antenna. However, it is worth mentioning that \gls{PZF} and \gls{OMP} require perfect knowledge of the \gls{CSI} and by increasing the number of antennas in \gls{BS} it reduces the spectral efficiency whereas our proposed methods only exploit \gls{RSSI} measurements which we kept  to $32$ and can therefore significantly improve the data rate of the system by reducing the signaling overhead. To clarify, in $N_{\sf{T}} = 128$ where the proposed method and \gls{PZF} have almost same performance, our proposed method requires $K \times N_{\sf{U}} \times N_{b} = 1024$ bits to feedback the \glspl{RSSI} where $N_{b} = 8$. On the other hand, \gls{PZF} in same configuration requires $N_{\sf{T}} \times N_{\sf{U}} \times 2 \times N_{b} = 4096$ to feedback the \gls{CSI} to the \gls{BS} where we find that $N_{b} = 4$ to keep the sum-rate same as fully precision. As a result, in same configuration and sum-rate performance, \gls{PZF} requires $4$ times more feedback bits in comparison with our proposed method.

\begin{figure}
    \centering
    \includegraphics[width=0.6\columnwidth]{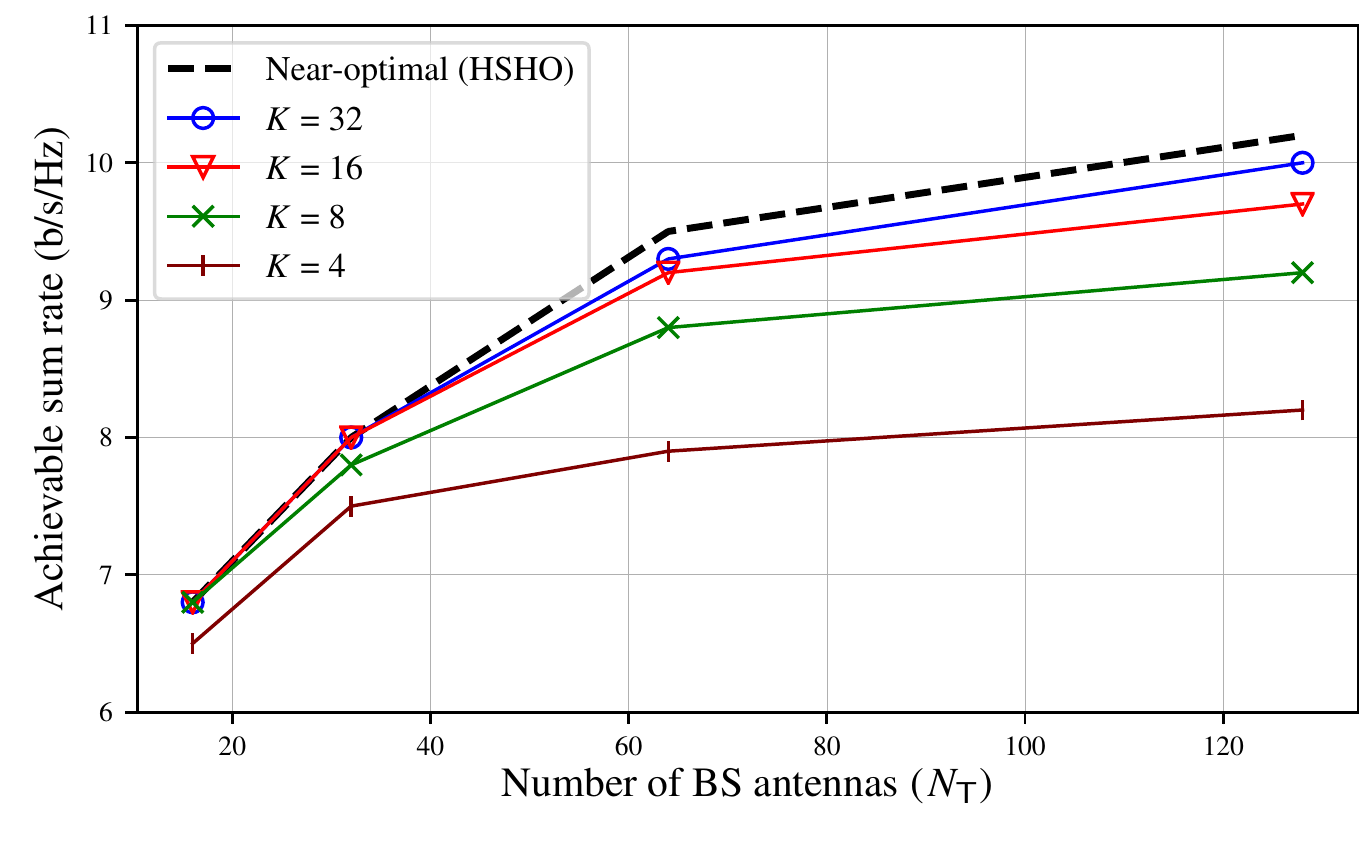}
    \caption{Sum-rate performance of HBF design in ``AFP-Net'' versus different number of antenna in BS ($N_{\sf{T}}$) in ``Extended area'' for different number of RSSI ($K$) ($\sigma^2= -130$~dBW, $N_{\sf{U}}=4$).}
    \label{fig:NT_K_test}
\end{figure}

Fig.~\ref{fig:NT_K_test} shows how the number $K$ of RSSI measurements affects the sum-rate performance when the number of antennas is increased. It can be seen that the required number of RSSIs to achieve near-optimal sum-rate depends on the number of antenna. For instance, with $N_{\sf{T}} = 16$ using $K=8$ or even $4$, the \gls{DNN} has similar performance as larger number of \gls{RSSI} such as $K=16$ or $32$. However, for larger number of antennas, the \gls{DNN} needs more information to design the \gls{HBF} and as a result more \glspl{RSSI} transmission are required. 

\section{Conclusion}\label{Conclusion}

Hybrid beamforming is an essential technology for massive MIMO systems that allows to reduce the number of RF chains and therefore increase the energy efficiency of the system. However, the design of digital and analog precoders is challenging, and the estimation of the CSI introduces important signaling overhead, especially in FDD communication.
To alleviate this issue, we instead proposed to rely on RSSI feedback, improving the spectral efficiency of the communication system. The \gls{SS} burst was efficiently designed so that the RSSIs provide maximum information about the \gls{CSI}. We then proposed to design the hybrid beamforming by using unsupervised deep-learning methods. This unsupervised learning approach leads to decreased training time and cost since the system can be trained using only channel measurements without the costly need to obtain optimal solutions. The deep-learning methods select the \gls{AP} from a codebook designed to reduce the complexity without sacrificing the sum-rate performance. 
Finally, the performance of the proposed algorithm was evaluated using the realistic \emph{deepMIMO} channel model. The results demonstrate that, despite not having access to the full CSI, the BS can be trained to robustly design the hybrid beamforming while achieving a similar sum-rate as the HSHO near-optimal hybrid precoder that was introduced as a benchmark. Moreover, the proposed method can be implemented in real-time systems thanks to its low computational complexity.


\bibliographystyle{IEEEtran}
%
\bibliography{bib/HBF_DL.bib}

\end{document}